\documentclass[preprint,12pt,3p]{elsarticle}

\usepackage{amssymb}
\usepackage{multirow}

\usepackage{doi}

\usepackage{caption} 
\usepackage{graphicx,subfigure}
\usepackage{verbatim}
\usepackage{makecell}
\usepackage{longtable}

\journal{Nucl. Instrum. Methods. Phys. Res. A}

\begin{document}

\begin{frontmatter}

\title{ Comparison of Plastic Antineutrino Detector Designs in the Context of Near Field Reactor Monitoring  }

\author[label1,label2]{Mustafa Kandemir}

\address[label1]{Department of Physics Eng., Istanbul Technical University, 34469, Istanbul, Turkey}
\address[label2]{Department of Physics, Recep Tayyip Erdogan University, 53100, Rize, Turkey\fnref{label4}}

\cortext[cor1]{corresponding author}

\ead{mustafa.kandemir@erdogan.edu.tr}


\author[label1]{Altan Cakir\corref{cor1}}
\ead{cakir@cern.ch}


\begin{abstract}

We compare existing segmented plastic antineutrino detectors with our new geometrically improved design for antineutrino detection and light collection efficiency. The purpose of this study is to determine the most suitable design style for remote reactor monitoring in the context of nuclear safeguards. Using Monte Carlo based GEANT4 simulation package, we perform detector simulation based on two prominent experiments: Plastic antineutrino detector array (Panda) and Core monitoring by reactor antineutrino detector (Cormorad). In addition to these two well-known designs, another concept, the Panda2, can be obtained by making a small variation of Panda detector, is also considered in the simulation. The results show that the light collection efficiency of the Cormorad is substantially less with respect to the other two detectors while the highest antineutrino detection efficiency is achieved with the Cormorad and Panda2. Furthermore, as an alternative to these design choices, which are composed of an array of identical rectangular-shaped modules, we propose to combine regular hexagonal-shaped modules which minimizes the surface area of the whole detector and consequently reduces the number of optical readout channels considerably. With this approach, it is possible to obtain a detector configuration with a slightly higher detection efficiency with respect to the Panda design and a better energy resolution detector compared to the Cormorad design.

\end{abstract}


\end{frontmatter}


\section{Introduction}
\label{sec1}

Since the first observation of reactor electron antineutrinos with an antineutrino detector in 1956 \cite{Cowan}, numerous experiments have been carried out around the world to understand the properties of the antineutrinos as an elementary particle. The knowledge obtained from these studies revealed the idea that the core of a nuclear reactor could be monitored by using antineutrinos. \cite{Mikaelyan}. The pioneering Rovno \cite{rovno} and the following SONGS \cite{Bernstein1} experiment have successfully demonstrated the accuracy of this idea by showing an explicit correlation between the change in antineutrino spectrum and the evolution of the reactor fuel isotopes.  

\begin{figure}[!htb]
\centering
 \includegraphics[width=0.50\linewidth]{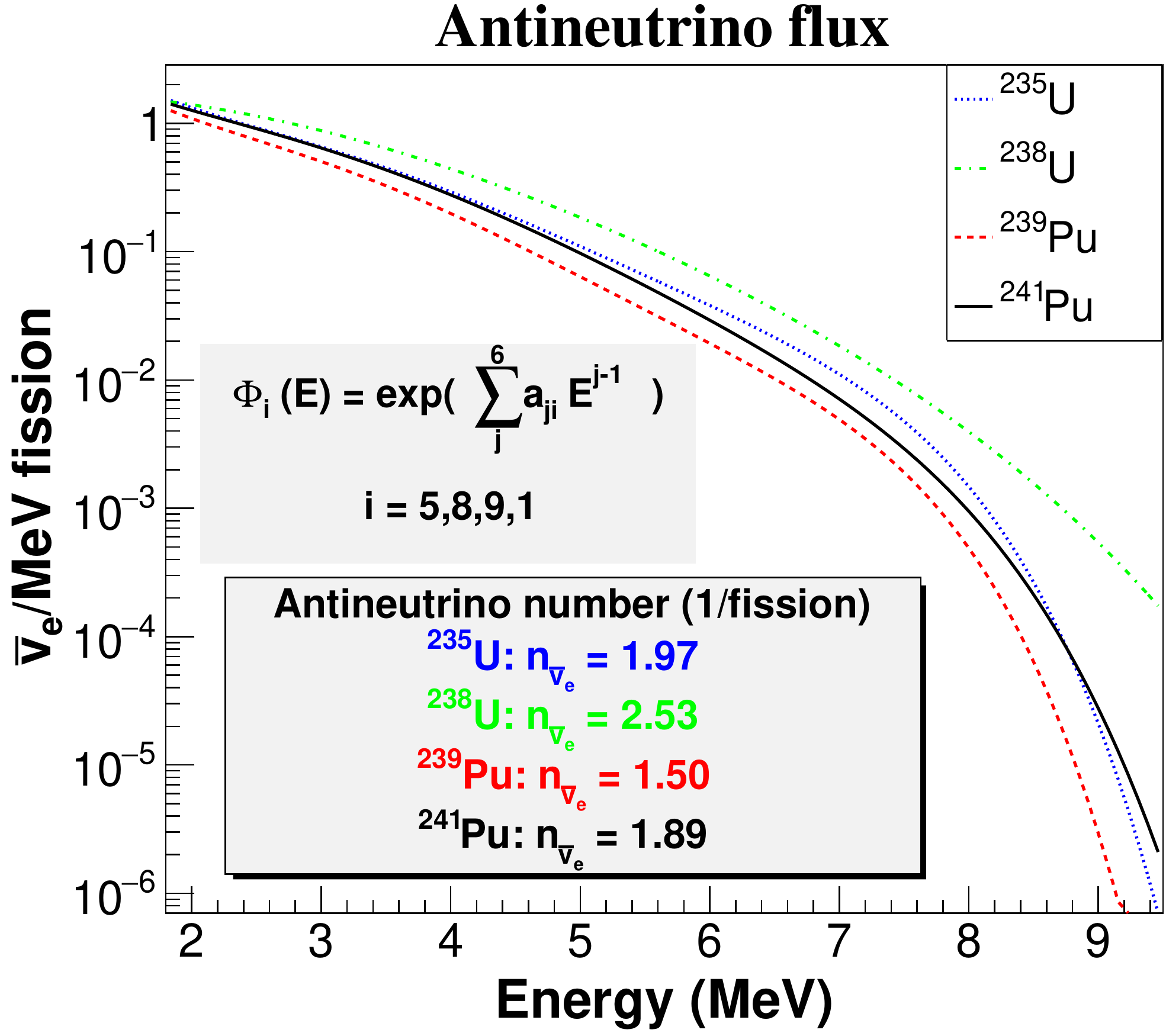}
\caption{The coefficients $a_{ji}$ are taken from Huber \cite{Huber} and Mueller \cite{Muller} study. Integrating over energy (from 1.8 to 10 MeV) gives the number of emitted antineutrinos per fission.}
\label{fig:fig1}
\end{figure}

Antineutrino emission in nuclear reactors mainly arises from the fission of four main fuel isotopes: $^{235}U$, $^{239}Pu$, $^{241}Pu$ and $^{238}U$. When one of these isotopes is exposed to fission, the resulting unstable fission products undergo beta decay and emit antineutrinos ( $ {^{A}_{Z}X} \rightarrow  {^{A}_{Z+1}Y} + e^- + \overline{\nu}_e$). The energy of the emitted antineutrinos can be inferred by two different methods. The first method is theoretical summation approach, where the antineutrino spectrum associated with one of the four fuel isotopes is computed as the sum of the contributions from all fission products. The second is an experimental method in which thin targets of foils of $^{239}Pu$, $^{241}Pu$, $^{235}U$ and $^{238}U$ are irradiated with neutrons to induce fission. The spectra of the electrons emitted by the beta decay products of fission isotopes are measured and then converted into antineutrino energy spectra. With the help of both these methods, Huber and Mueller provide a parameterization for the reactor antineutrino emission spectrum in a sixth-order polynomial form for each fissioning isotopes. Fig. $\ref{fig:fig1}$ shows the emitted antineutrino energy spectrum per fission of each fissioning isotope drawn by the recent Huber \cite{Huber} and Mueller \cite{Muller} model.

\begin{figure}[!htb]
    \centering
    \subfigure[ Time evolution of fission fraction of each isotope over the fuel cycle (from Ref. \cite{Bemporad}) ]
    {
        \includegraphics[width=0.45\linewidth]{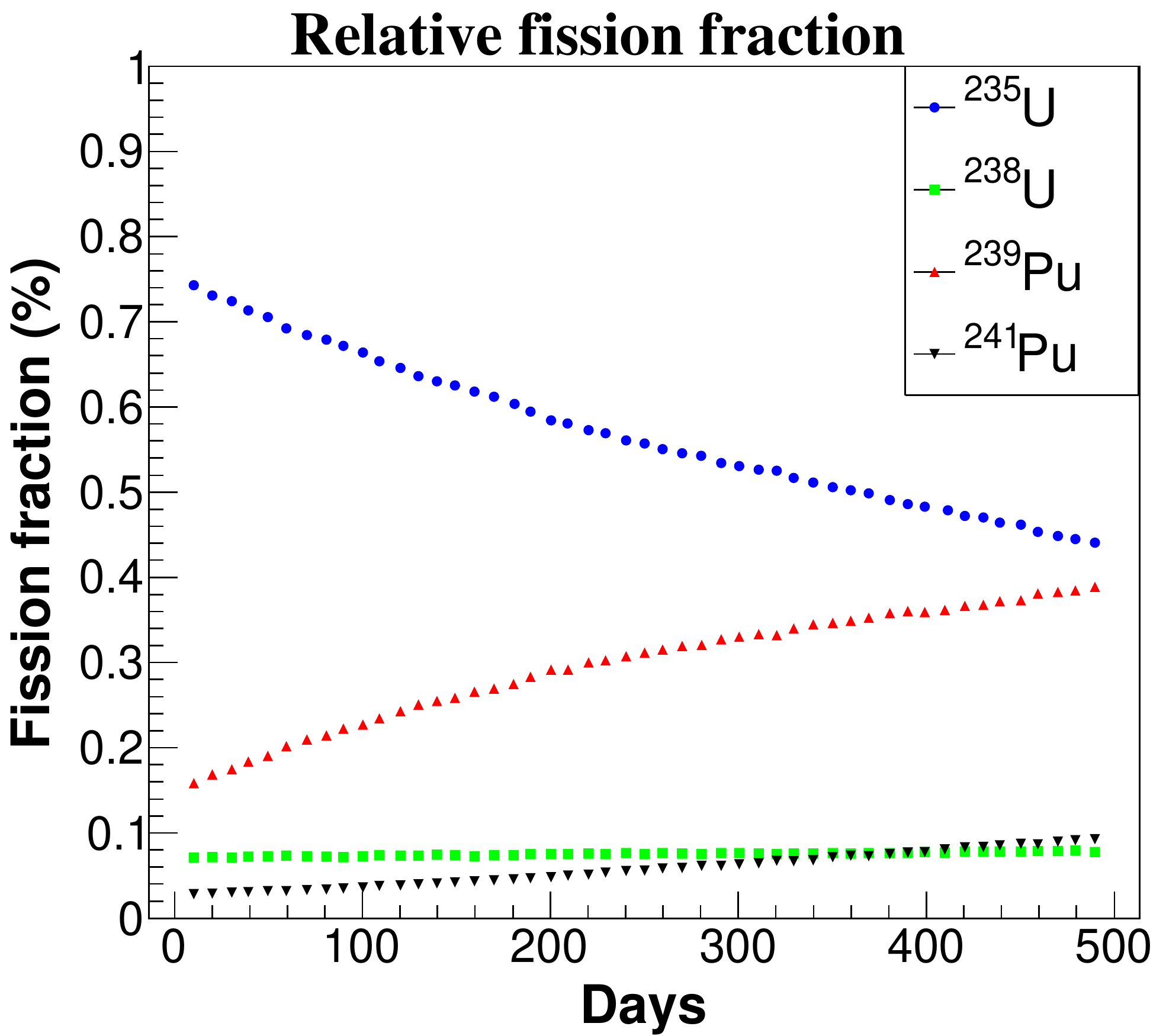}
        \label{fig:fig2a}
    }
    \quad
    \subfigure[ Expected antineutrino energy spectrum at the beginning, middle and end of the reactor fuel cycle. ]
    {
       \includegraphics[width=0.45\linewidth]{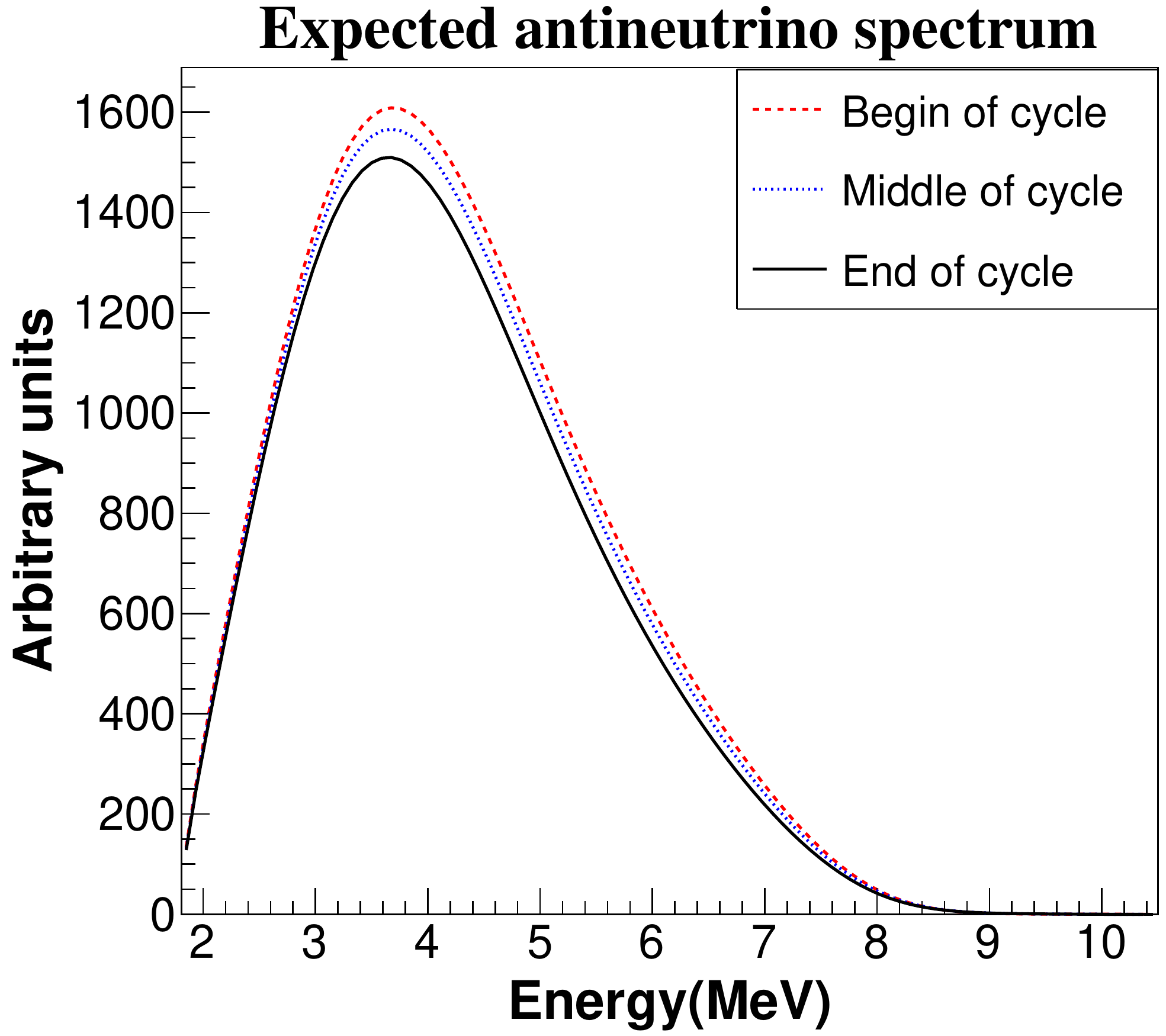}
        \label{fig:fig2b}
    }
    \caption{ As the reactor fuel burns up, the contribution of each isotope to fission changes. Since the fission of each isotope produces a unique antineutrino spectrum, this leads to a shift in the detected antineutrino energy spectrum.   }
    \label{fig:fig2}
\end{figure}

Since the emitted energy spectra per fission vary according to fuel isotopes (Fig. $\ref{fig:fig1}$) and the relative contribution to fission of these isotopes evolves over the reactor fuel cycle (Fig. $\ref{fig:fig2a}$), measurement of the change in antineutrino energy spectrum over the course of fuel cycle gives information regarding fissile content of the reactor core. Fig. $\ref{fig:fig2b}$ shows the state of the expected antineutrino energy spectrum at the beginning, middle and end of the reactor fuel cycle.  
  
\begin{figure}[!htb]
    \centering
    \subfigure[ Schematic representation of antineutrino detection event with hexagonally shaped detector.  ]
    {
        \includegraphics[width=0.45\linewidth]{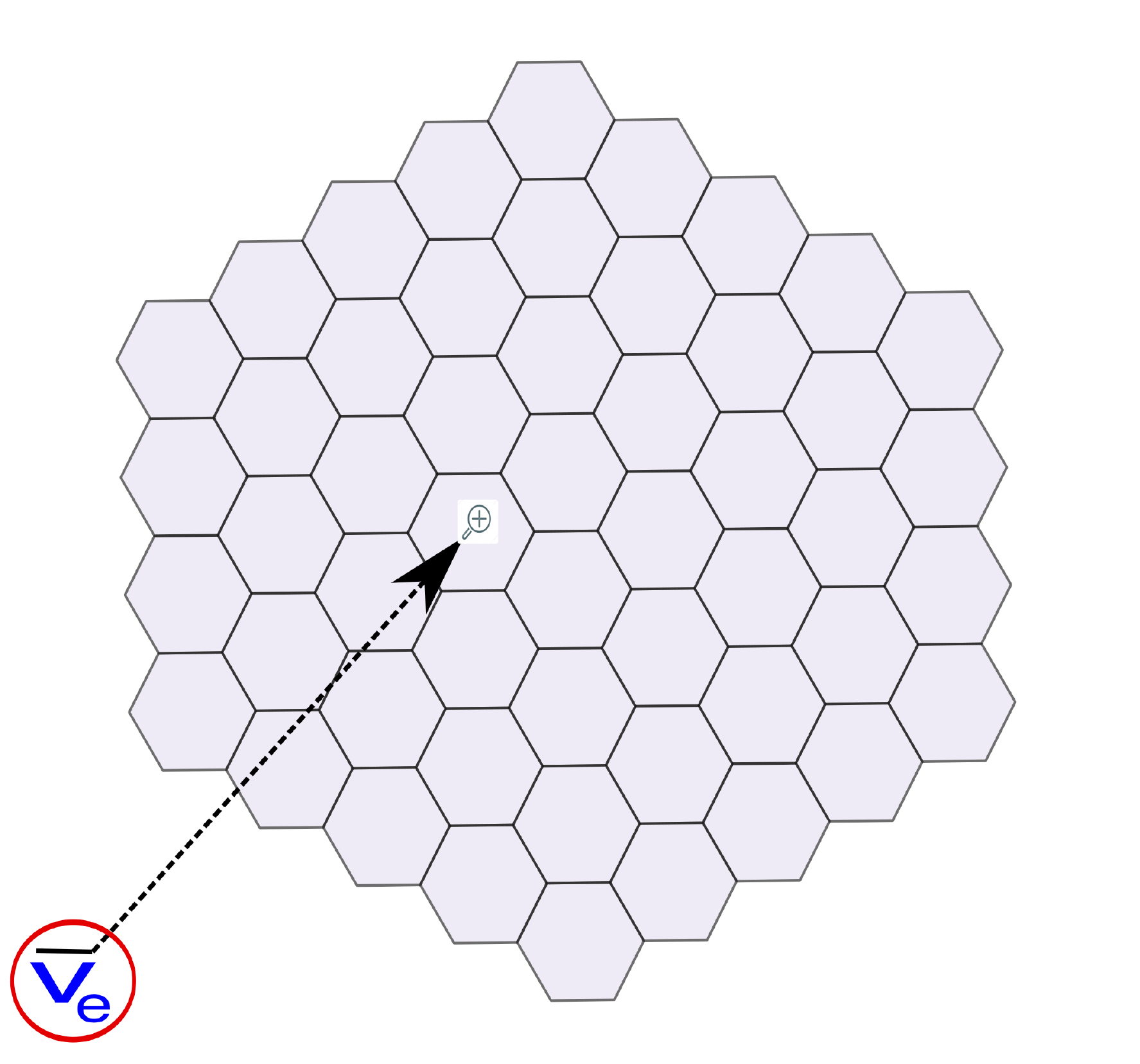}
        \label{fig:fig3a}
    }
    \quad
    \subfigure[A close-up view of the antineutrino detection event shown on the left.    ]
    {
       \includegraphics[width=0.45\linewidth]{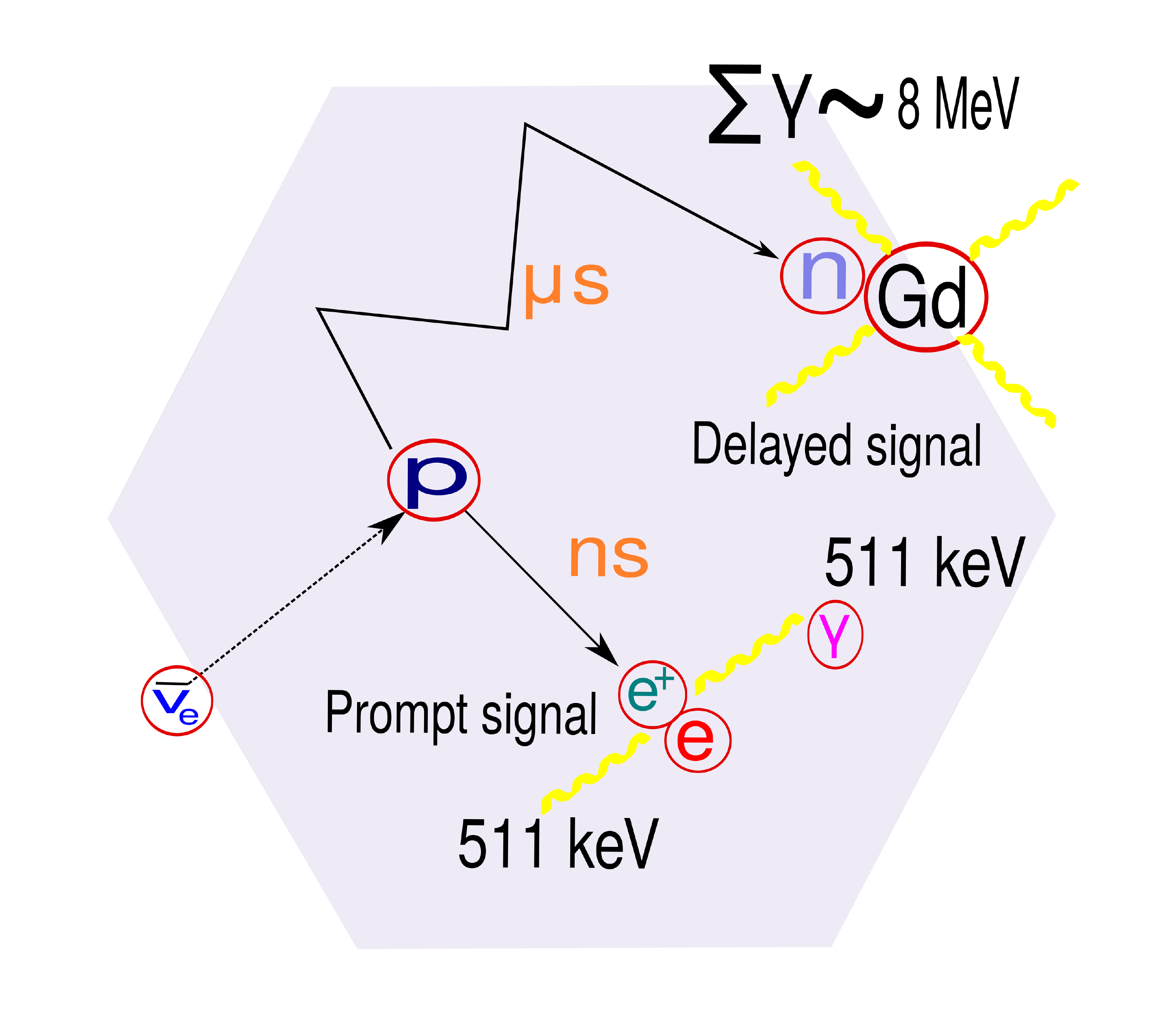}
        \label{fig:fig3b}
    }
    \caption{ Inverse beta decay event. Gadolinium is used as a neutron capture agent to enhance detector's sensitivity to antineutrinos. }
    \label{fig:fig3}
\end{figure}

Reactor electron antineutrinos are detected through the inverse beta decay process $\overline{\nu}+p \rightarrow e^+ + n $ ($E_{\overline{\nu}}>1.8$ MeV), which is the largest cross-section for antineutrinos of few MeV energy. When antineutrino interacts via the weak force with a quasi-free proton, a positron and a neutron arise. In a scintillator-based detector, the positron deposits its energy via ionization and when at rest it annihilates with an electron producing two 0.511 MeV gamma photons traveling in opposite directions. The two gamma interact with the crossed material by Compton scattering indirectly releasing energy by ionization. These processes occur immediately (within a few ns) and it is called the prompt signal. On the other hand, neutron losses its energy via elastic scattering and eventually gets captured by a hydrogen nucleus or a neutron capture agent added to the scintillator to enhance neutron capture efficiency. Depending on the capture agent, its location, and concentration in the scintillator volume, this process can take from one to hundreds of microseconds, and it is called the delayed signal. The correlated time difference between the prompt signal and the delayed signal tags antineutrino event and stands out strongly against background. Fig. $\ref{fig:fig3}$ shows schematic representation of inverse beta decay event.

The main challenge of antineutrino detection is selecting the signal and distinguishing it from the background, which is very high compared to antineutrino event. In a near-field measurement, backgrounds present close to a nuclear reactor can be classified in two parts as correlated and uncorrelated. The uncorrelated ones consist of two independent events (i.e., a prompt-like and a neutron-like event) that randomly occur within the prompt-delayed time interval. The prompt-like and neutron-like signals are mainly created by reactor induced gamma and low energy neutrons. On the other hand, correlated backgrounds consist of a single event that produces two time correlated signal similar to antineutrino signature. This type of backgrounds is mostly related
to cosmogenic muons/neutrons and spallation neutrons. In a segmented plastic antineutrino detector, antineutrino events can be discriminated from backgrounds by using hit pattern identification and energy deposition profile.

In the application, a modest-sized detector placed at a few meters distances from a typical thermal power reactor can detect reactor antineutrinos from outside the containment building. The expected number of detected antineutrinos per unit time and energy depend on the emitted antineutrino energy spectrum per fission of each of the four isotopes ($\Phi_i (E_{\overline{\nu}}$), thermal energy release per fission of each istope in the reactor core ($E_i$), antineutrino-proton interaction cross section ($\sigma (E_{\overline{\nu}})$), distance between the detector and the center of the reactor core (R), thermal power of the reactor (P), free proton number ($N_p$) in the volume of the detector, detection efficiency of the detector ($\epsilon$) and fission fraction of each isotope at time t ($\alpha_i(t)$).

\begin{eqnarray}
\label{eq:1}
\frac{dN(E_{\overline{\nu}},t)}{dE_{\overline{\nu}}dt} =\frac{1}{4 \pi R^2 }  N_p  \epsilon \frac{P}{\sum_{i=5,8,9,1} \alpha_i(t) E_i}  \sum_{i=5,8,9,1} \alpha_i(t)  \Phi_i (E_{\overline{\nu}}) \sigma (E_{\overline{\nu}}) 
\end{eqnarray}

Integrating over energy (1.8 to 10 MeV) gives the total number of detected antineutrinos per unit time. For a measurement time interval T, where the fission fraction of each isotope is almost constant, the equation becomes:

\begin{eqnarray}
\label{eq:2}
n_{\overline{\nu}} =\frac{1}{4 \pi R^2 }  N_p \epsilon T \frac{P}{<E_f>}  \sum_{i=5,8,9,1} \alpha_i  \sigma_i 
\end{eqnarray}

In the above equation, $\sigma_i$ is the cross section per fission of isotope i and $<E_f>$ is the average energy release per fission. 

\begin{eqnarray}
\label{eq:3}
\sigma_i =\int_{1.8}^{10} \Phi_i (E_{\overline{\nu}}) \sigma (E_{\overline{\nu}})  dE_{\overline{\nu}} \quad (cm^2/fission), \quad <E_f> = \sum_{i=5,8,9,1} \alpha_i(t) E_i 
\end{eqnarray}

Using the naive cross section of inverse beta decay \cite{Vogel2} and Huber-Mueller flux model, cross section per fission of four isotopes is calculated. Energy release per fission of each isotope is obtained from ref \cite{Erpf}. The results are shown in Table $\ref{table:table1}$.

\begin{table}[!htb]
    \center 
    \caption{ The calculated cross section per fission value of each isotope.  }
 \label{table:table1}
 \begin{tabular}{ |p{1.5cm}|p{4.2cm}|p{3.9cm}| }
\hline
Isotopes  & Cross section per fission ($\sigma_i,10^{-43}cm^2 /fission $) & Energy release per fission (MeV/fission) \\  
\hline
$^{235}U$ &   6.69  & 202.36 \\
\hline
$^{238}U$ &   10.18 & 205.99 \\
\hline
$^{239}Pu$ &  4.37 & 211.12 \\
\hline
$^{241}Pu$ &  6.04  & 214.26 \\
\hline
\end{tabular}
\centering
\label {cspf}
\end{table}

Although it is possible to monitor the operational status, thermal power and fissile content of a reactor core with modest-sized detectors, the sensitivity of the detectors to diversion of fissile material are not at the desired level for near field safeguard application \footnote{A perfect antineutrino detector should be able to detect 8 kg of plutonium diversion within 3 months and 75 kg of $^{235}U$ (LEU) diversion within 1 year. However, in the current technology, detectors are sensitive at the one sigma level to changes of roughly 50 kg of plutonium and 300 kg of Uranium in reactor fuel within the same time criterion. These values can be improved with the development of antineutrino detectors and data analysis methods \cite{Bernstein3}. }. In order for a detector to be used extensively worldwide in the field of non-proliferation as an additional tool to conventional method, it should meet some requirements of International Atomic Energy Agency (IAEA). These are improvement in the sensitivity of the detectors to diversion of fissile material, ease of deployment, safely operation, modest size, non-intrusive monitoring capability, preferably movable, ability to operate at sea level and low cost. Many different types of detectors have been proposed to achieve these goals. Recent years, plastic scintillation detectors are one of the most effective tools to improve for near-field safeguard applications. However, the overall efficiency and energy resolution of these detectors are under development. Although there are many applications for plastic scintillators, the overal efficiency and energy resolutions should be improved in terms of various detection techniques \cite{Bernstein3}. Therefore, we are going to use Geant4 \citep{Agostinelli} simulation package to search for the best suitable design option in terms of antineutrino detection and energy resolution efficiency by considering the most effective design for future of near-field safeguard detectors. 

The simulations are carried out in the following orders: First, we investigate the influence of some important design parameters on the neutron capture time and the neutron capture efficiency. Second, we compare the response of detectors when simple energy and time cuts are applied. Third, the light collection and detection efficiency of each detector are calculated in the case of using two different PMTs. Finally, the energy resolution of each detector is obtained and the positive and negative aspects of each detector design are discussed.


\section{Geometry and material description }

In this section, we describe a few existing detector designs such as Panda \cite{Oguri1, Oguri2}, Panda2 \cite{Kashyap}\footnote{Actually, this design is called Plastic quad in ref. \cite{Kashyap}. However, we call this design Panda2 since it is obtained by making a small change in Panda design. Furthermore, this detector is only proposed. It is not constructed. }, Ismran \cite{Mulmule}\footnote{Although Panda and Ismran (Indian scintillator matrix for reactor antineutrinos) are different projects, they use the same detector geometry.}, and Cormorad \cite{Battaglieri} along with our proposed design. The Panda/Ismran design consists of 100 identical modules and each module has a plastic scintillator bar of dimension 10cm$\times$10cm$\times$100cm, two acrylic cubic light guides of 10cm$\times$10cm$\times$10cm and two photomultipliers (PMTs). Light guides are connected to both ends of the plastic bar and then coupled to PMTs from both sides by means of optical cement (Fig. $\ref{fig:fig5a}$ up). On the other hand, Panda2 and Cormorad design are composed of 100 and 49 identical modules respectively. However, unlike the Panda, each module has four plastic bars instead of one. These four bars are combined together and then glued to the same light guide (Fig. $\ref{fig:fig5a}$ down). The dimension of each plastic bar in the Panda2 design is 5cm$\times$5cm$\times$100cm, and in the Cormorad design, 7cm$\times$7cm$\times$130cm.

In contrast to conventional designs, we use hexagonal plastic bars and light guides to form the detector \footnote {Actually, there are three possible module shapes that can be put together without leaving gaps: square, triangular and hexagonal prism. We choose hexagonal prism since it requires a minimum surface area to enclose the same detector volume.}. A module of our design includes a plastic bar with a height of 120 cm and a side length of 6 cm, two light guides with a height of 10 cm and a side length of 6 cm, and two PMTs. An array of these modules constitutes the whole detector. Hexagonal modules could be packed in two different ways: Rectangular shaped packing (RSP) and Hexagonal shaped packing (HSP). The type of packing determines the number of modules (93 for RSP and 91 for HSP ). Choosing hexagonal bars minimizes the total surface area of all the modules and thus reduces the number of PMTs required to readout a given detector volume. Fig. $\ref{fig:fig6}$ shows two different layouts of our proposed antineutrino detector design.

\begin{figure}[!htb]
\centering
\includegraphics[width=0.70\linewidth]{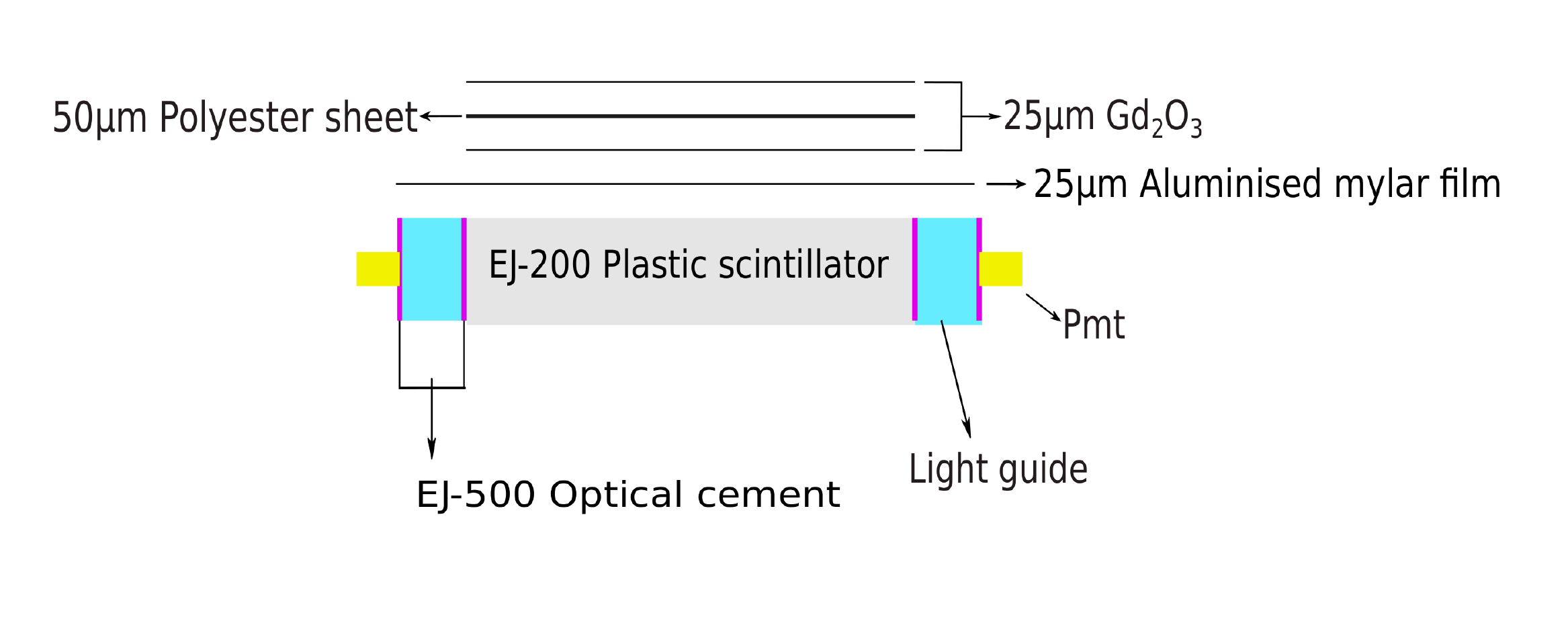}
\caption{ An antineutrino detector module. Each plastic bar is wrapped in the same way. }
\label{fig:fig4}
\end{figure}

\begin{figure}[!htb]
    \centering
    \subfigure[Panda module (up). Panda2 and Cormorad module (down). ]
    {
        \includegraphics[width=0.47\linewidth]{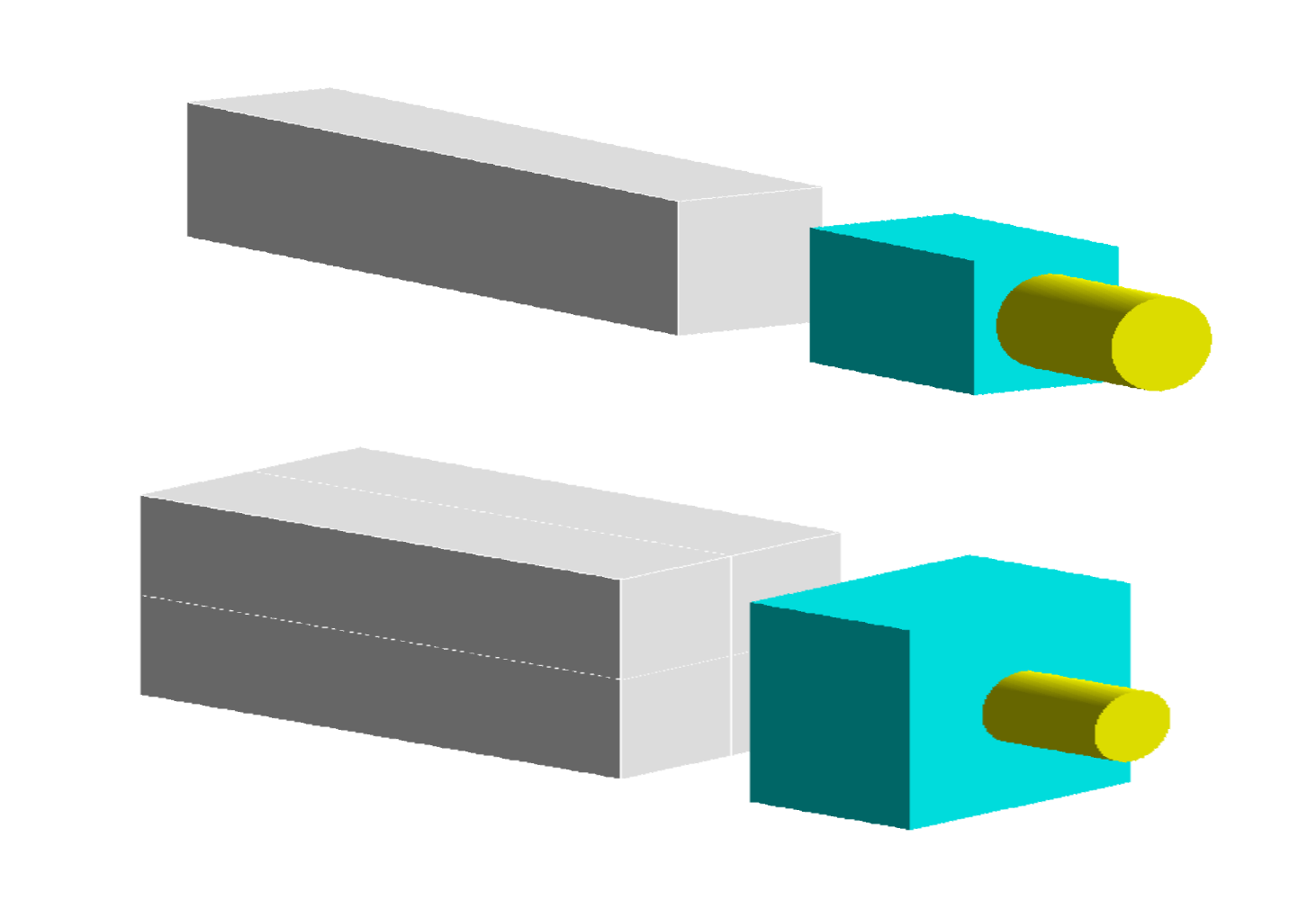}
        \label{fig:fig5a}
    }
    \subfigure[Panda detector.]
    {
       \includegraphics[width=0.47\linewidth]{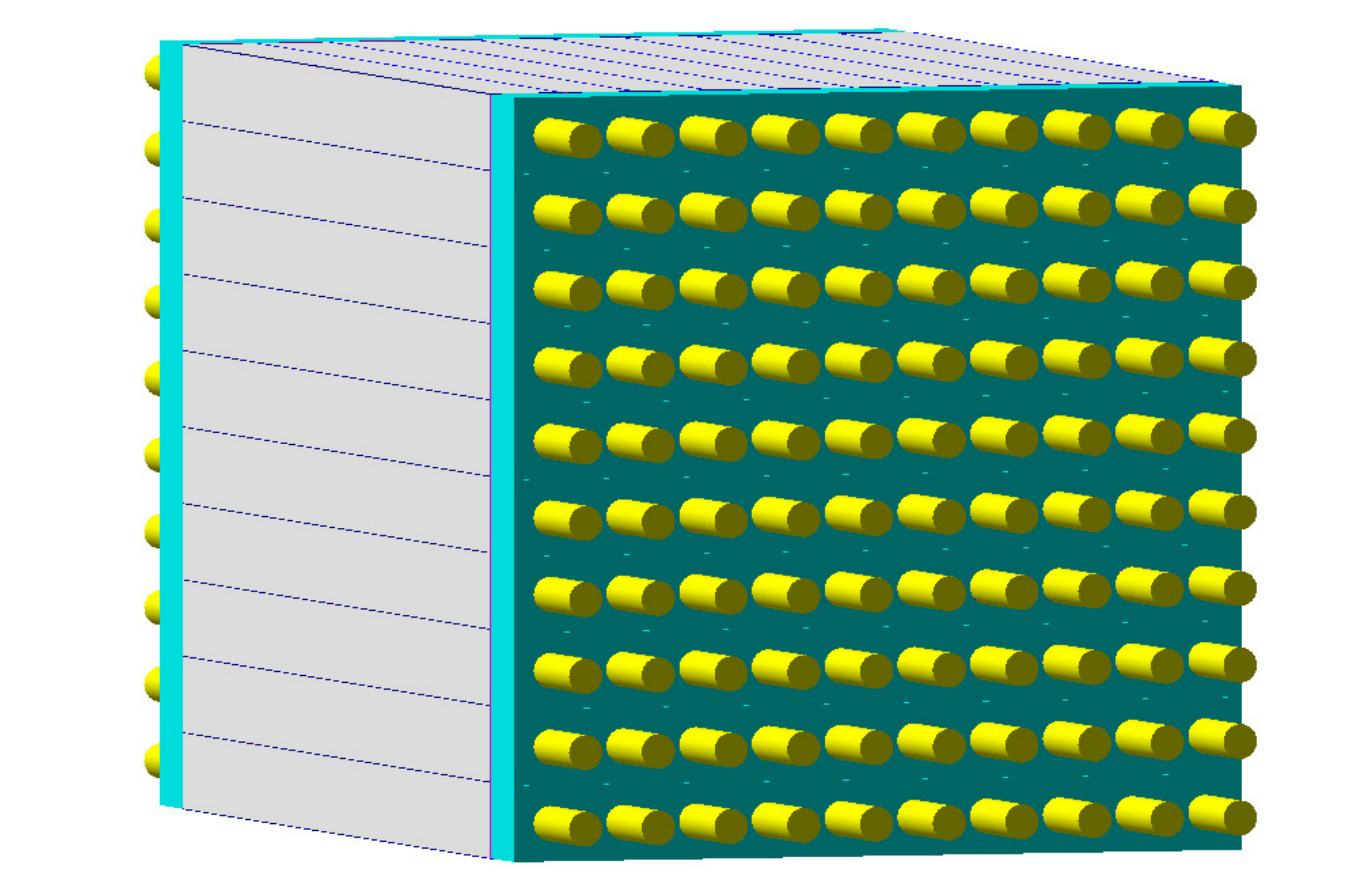}
        \label{fig:fig5b}
    }
    \caption{ Rectangular prism-shaped antineutrino detector modules are packed both rectangular (Cormorad) and cubic form (Panda and Panda2).  }
    \label{fig:fig5}
\end{figure}

\begin{figure}[!htb]
    \centering
    \subfigure[Rectangular shaped packing (RSP)]
    {
        \includegraphics[width=0.47\linewidth]{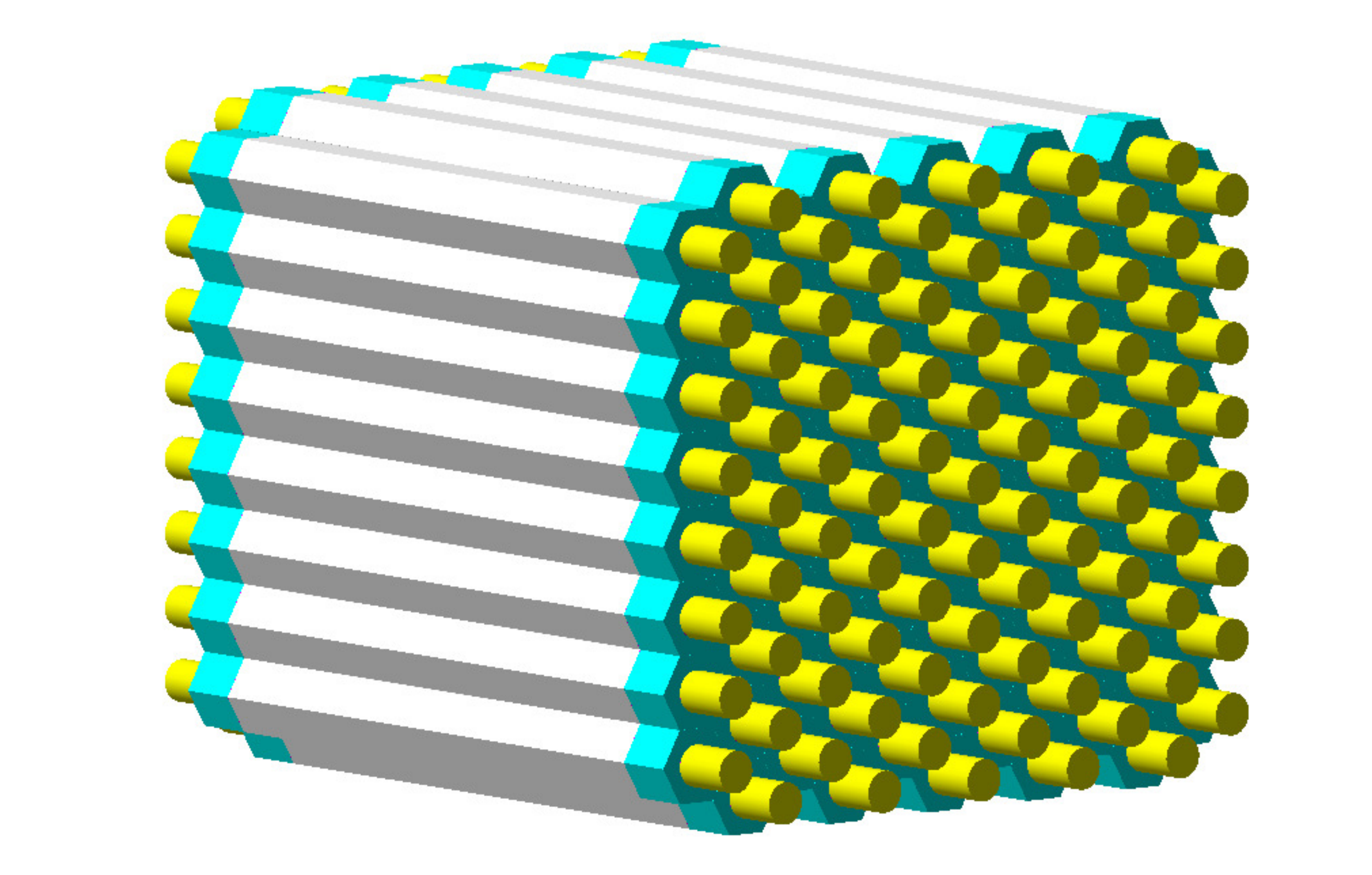}
        \label{fig:fig6a}
    }
    \subfigure[Hexagonal shaped packing (HSP)]
    {
       \includegraphics[width=0.47\linewidth]{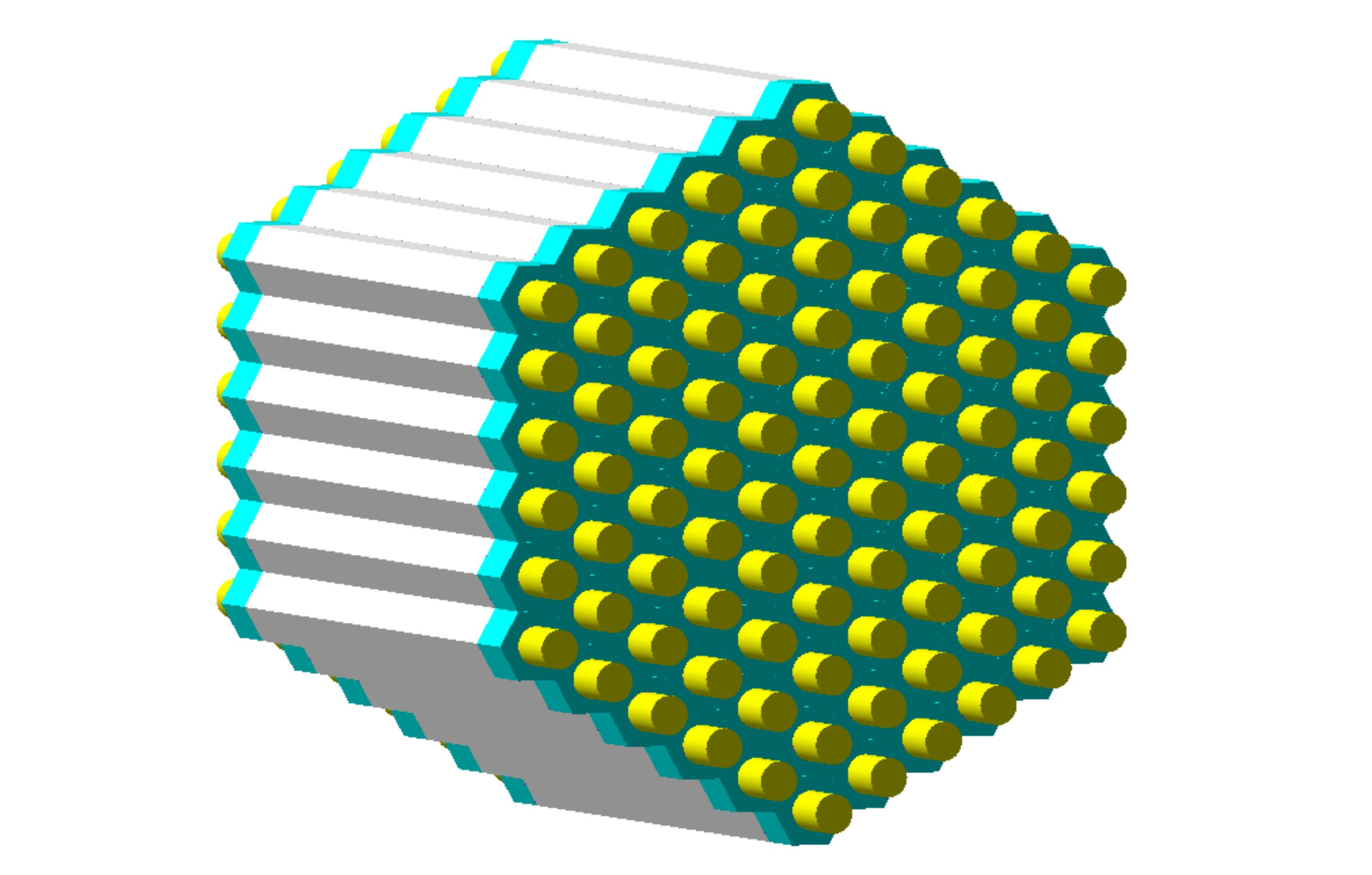}
        \label{fig:fig6b}
    }
    \caption{ Regular hexagonal prism-shaped antineutrino detector modules are packed both rectangular and hexagonal form. }
    \label{fig:fig6}
\end{figure}

The used plastic scintillator material is EJ-200 (ELJEN Technology \cite{Eljen}) and the optical cement is EJ-500. Two different PMT models are selected for light detection: 2-inch H6410 Hamamatsu PMT \cite{Hamamatsu} and 3-inch 9265B ET Enterprise PMT \cite{Et}. Each individual plastic scintillator bar and light guide are wrapped with 25 $\mu$m thick aluminized Mylar film to reflect scintillation photons. Additionally, each plastic bar is wrapped with gadolinium (Gd) coated polyester sheet to reduce neutron capture time. The sheet is made of three thin layers. A 50 $\mu$m thick polyester film is sandwiched between the two layers of 25 $\mu$m thick Gd$_2$O$_3$ coating. The sheet contains 4.9 mg of Gd per cm$^2$. The total Gd concentration (w/w) of Panda, Panda2, Cormorad, HSP and RSP are $0.19\%$, $0.38\%$, $0.27\%$, $0.18\%$ and $0.18\%$, respectively. The Gd concentration of the detectors changes with the thickness of Gd$_2$O$_3$ coating. The materials used and the wrapping method of the plastic bars are the same for each design and they are described as in the Panda experiment. Fig. $\ref{fig:fig4}$ shows a detailed view of the module content. 

Table $\ref{table:table2}$ describes the properties of the detectors used in the simulation. The necessary parameters to perform optical photon simulation such as emission spectrum of the scintillator, refractive index of the materials, quantum efficiency of the PMT photocathode and the property of the reflector are available in our previous study \cite{musti}.\footnote{In our previous work, we study detail simulation of optical photon transportation and detection with plastic antineutrino detector modules. The parameters affecting the light collection efficiency and the energy resolution of the detectors are examined in detail in that study \cite{musti}. }

\begin{table}[!htb] 

  \center 
 \caption{ The properties of the detectors used in the simulation. The variable x, y and z represent the dimensions of the plastic bars. For hexagonal bars, x indicates the edge length of the hexagons. The z size of the light guide is 10 cm for each module. N$_m$ and N$_{pmt}$ indicate the total number of modules and PMTs used in each detector type. N$_{p}$ indicates the total number of protons in the volume of the detectors. } 
\label{table:table2}

    \begin{tabular}{ |p{2.5cm} |p{3.2cm} | p{2.5cm} | p{3.2cm} | p{1.2cm} | p{1.2cm} | }

\hline
\multirow{2}{*}{ \shortstack[l]{ \textbf{Quantity} }  } & \multicolumn{3}{|c|}{ \textbf{Existing Design} } & \multicolumn{2}{|c|}{ \textbf{Proposed Design} } \\
\cline{2-6}

                   & \textbf{Panda \cite{Oguri1, Oguri2}- \newline Ismran \cite{Mulmule}} & \textbf{Panda2 \cite{Kashyap}}  & \textbf{Cormorad \cite{Battaglieri}}  &  \textbf{HSP}  &  \textbf{RSP}   \\
 \hline
 
  x (cm) &  10 & 5 & 7 & 6 & 6  \\
 \hline
 y (cm) &  10 & 5 & 7 & - & -  \\
 \hline
 z (cm)          &  100  & 100  & 130  & 120  & 120  \\
 \hline
 $N_m$           &  100  & 100  & 49   & 91   & 93  \\
 \hline
 $N_{Pmt}$       &  200  & 200  & 98   & 182  & 186  \\
 \hline
 Mass (kg)       &  1023 & 1023 & 1277 & 1045 & 1068 \\
 \hline
 Volume ($m^3$)  &  1.00 & 1.00 & 1.25 & 1.02 & 1.04 \\
 \hline
 $N_p$ $(x10^{28})$ &  5.17 & 5.17 & 6.45 & 5.28 & 5.40 \\
 \hline
  \raggedright Total Gd concentration $\%$(w/w) &  0.19 & 0.38 & 0.27 & 0.18 & 0.18 \\
 \hline

\end{tabular}
 
\end{table}

\section{Simulations}

\subsection{Inverse beta decay}

\begin{figure}[!htb]
    \centering
    {
        \includegraphics[width=0.47\linewidth]{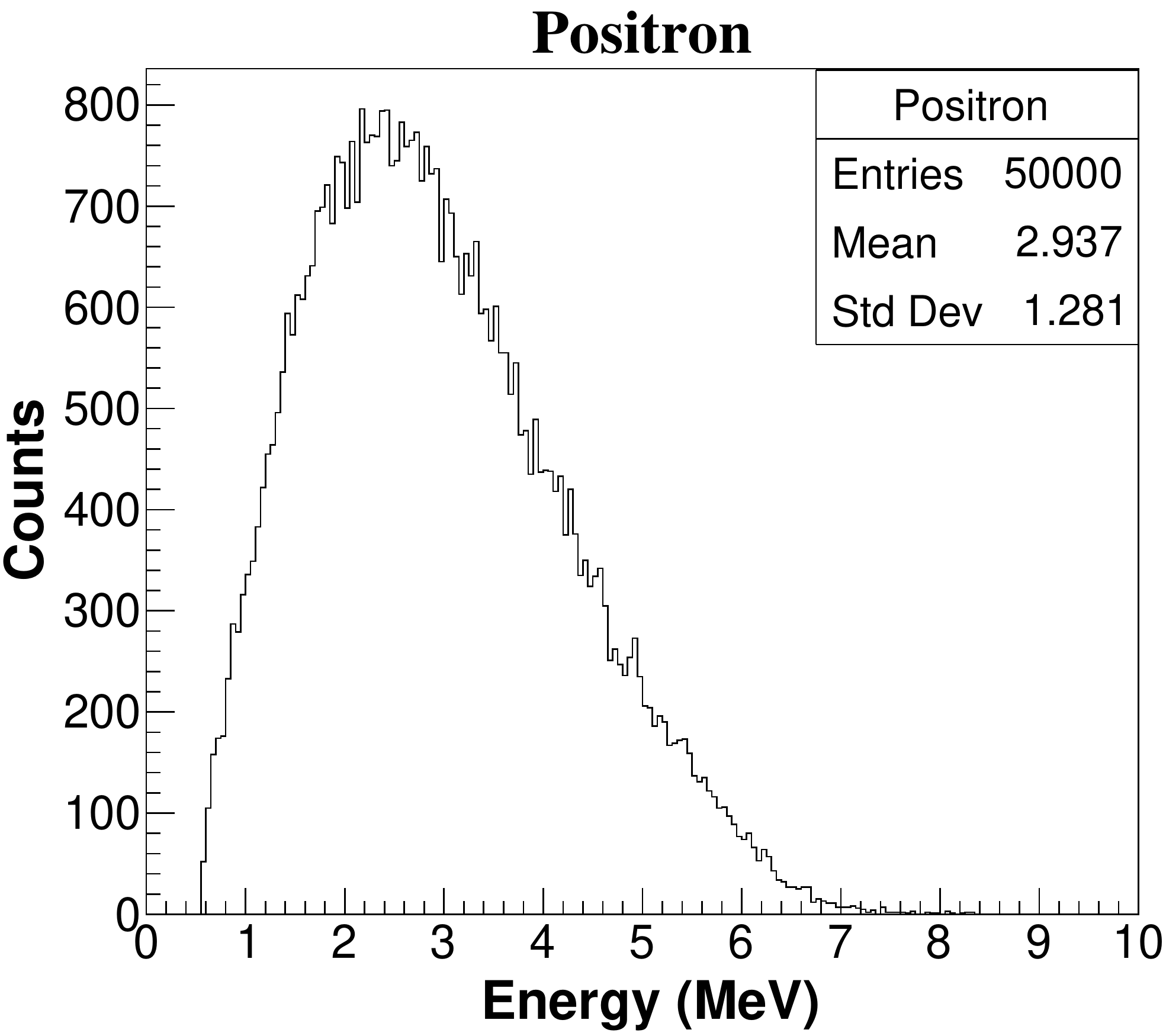}
    }
    {
       \includegraphics[width=0.47\linewidth]{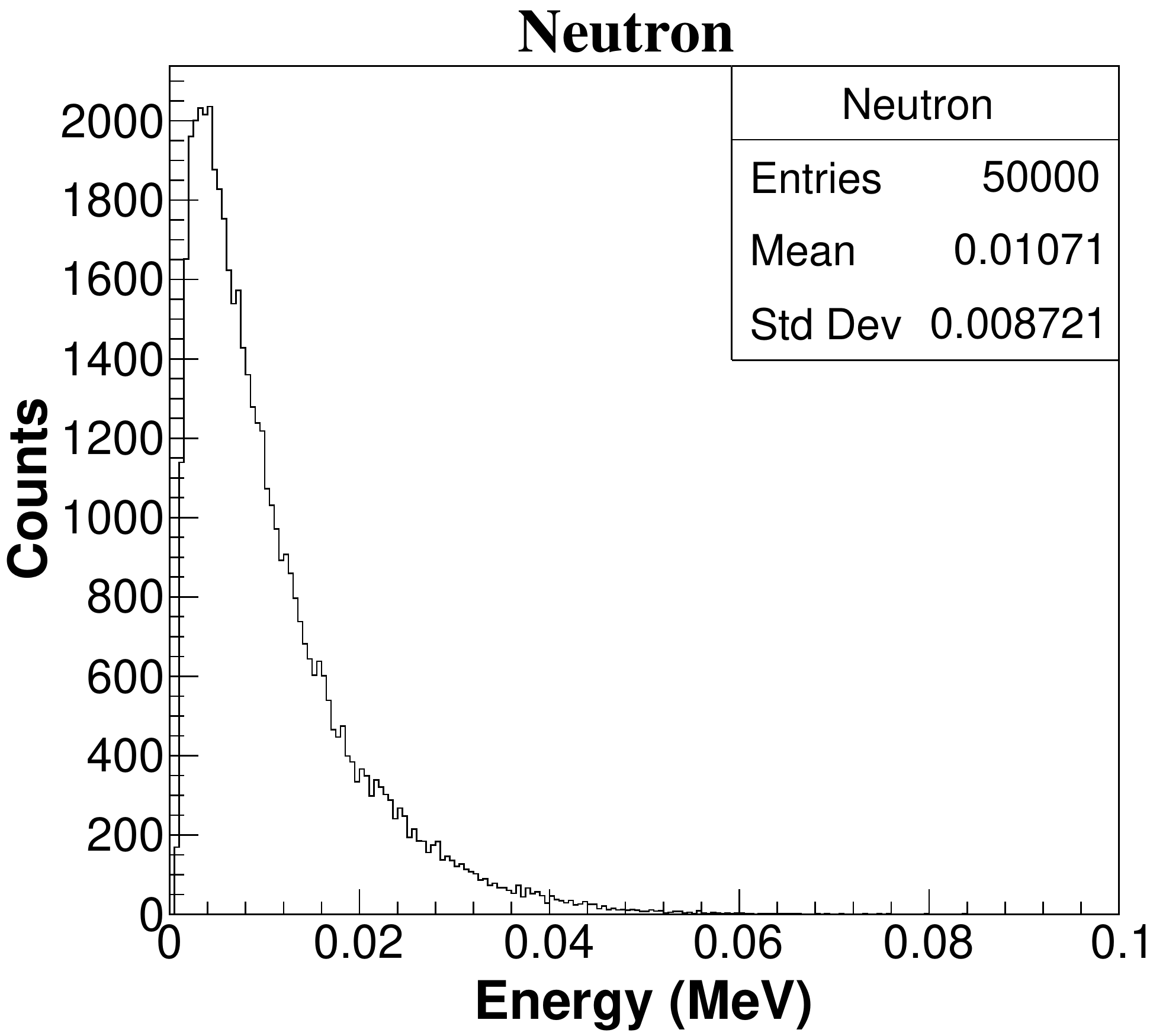}
    }
    \caption{ Products of inverse beta decay: positron and neutron.  }
    \label{fig:fig7}
\end{figure}

In order to compare our proposed plastic antineutrino detector's efficiency with the existing plastic antineutrino detectors such as Panda, Panda2, Ismran and Cormorad, inverse beta decay (IBD) event is simulated for each design option. Antineutrino interaction position with the proton is generated randomly in a point inside the detector active volume. The energy of the antineutrino is taken from the expected antineutrino energy spectrum for each event (see Fig.$\ref{fig:fig2b}$). The energy of the positron and neutron, which depends on the antineutrino energy, is derived from the inverse beta decay kinematics \cite{Vogel}. Fig. $\ref{fig:fig7}$ shows the initial energy distribution of the positron and neutron created as a primary particle in the simulation.

These two primaries and subsequently produced secondaries are tracked throughout the detector volume until they disappear. The total deposited energy within the detector due to the original particle positron and neutron, the neutron capture time, its capture position, and the identity of nuclei which capture neutron are recorded in each event for later analysis\footnote{The parameters neutron capture time, neutron capture position and the identity of nuclei which capture neutron are not physically observable and their use is limited to a better understanding of the interaction dynamics rather than helping data analysis.}. All the simulation samples are carried out with the same number of events.

\subsubsection{Neutron capture efficiency and time}  

\begin{figure}[!htb]
    \centering
    \subfigure[Neutron capture efficiency of gadolinium, hydrogen and the sum of the two. The results are shown for the HSP design.]
    {
        \includegraphics[width=0.47\linewidth]{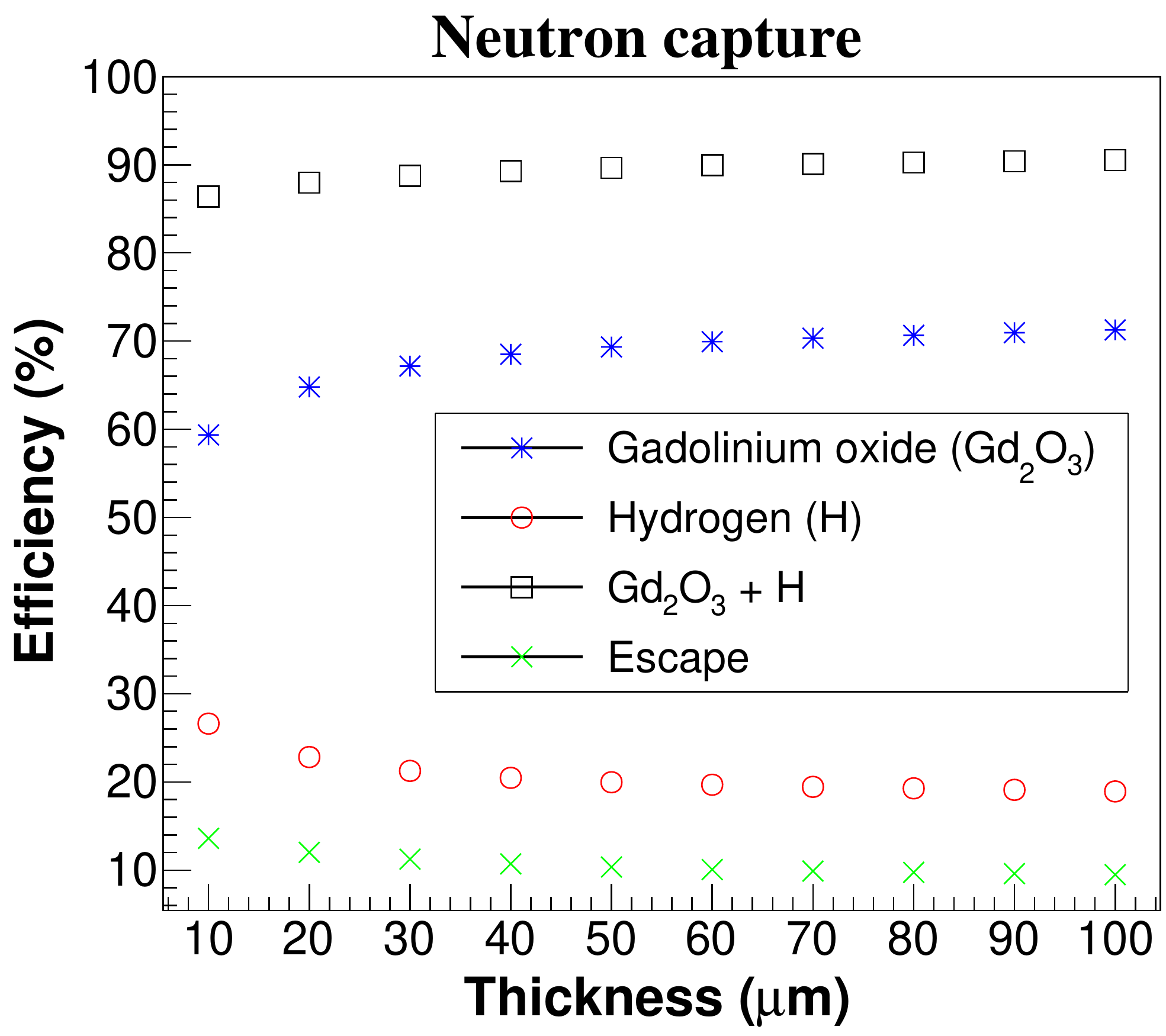}
        \label{fig:fig8a}
    }
    \subfigure[ Mean neutron capture time as a function the Gd$_2$O$_3$ layers thickness.]
    {
       \includegraphics[width=0.47\linewidth]{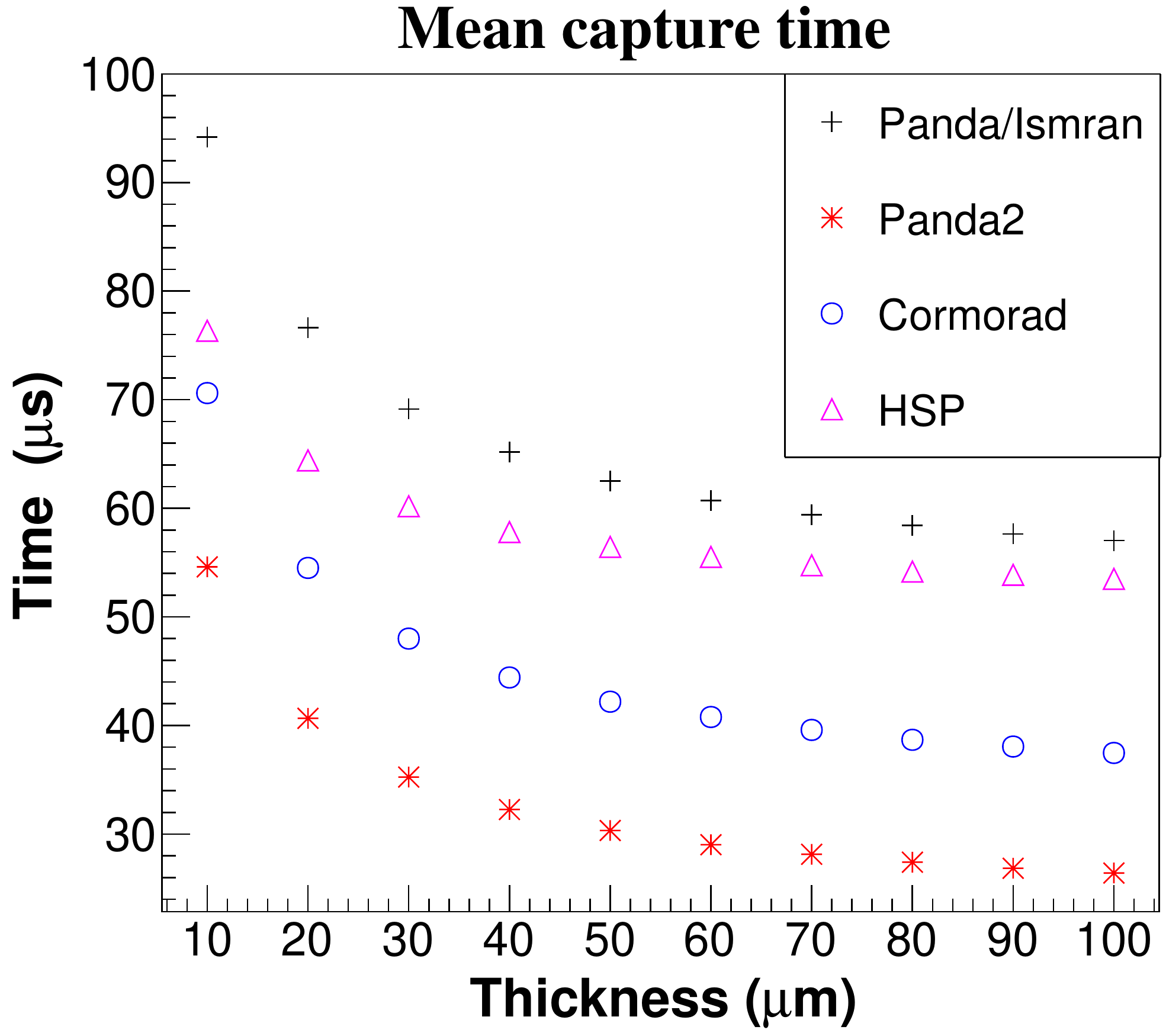}
        \label{fig:fig8b}
    }
    \caption{The impact of Gd$_2$O$_3$ layer thickness on the neutron capture efficiency and the neutron capture time.  }
    \label{fig:fig8}
\end{figure}

We first investigate the effect of Gd-oxide layer thickness on the neutron capture efficiency and the neutron capture time for each detector design. The IBD events are simulated for different thickness of Gd$_2$O$_3$ layers. Fig. $\ref{fig:fig8}$ shows the effect of thickness on the neutron capture efficiency and the neutron capture time. From Fig. $\ref{fig:fig8}$ we see that an increase in Gd$_2$O$_3$ layer thickness provides dual benefits; it increases the overall capture efficiency of neutron (Gd$_2$O$_3$ + H in Fig. $\ref{fig:fig8a}$) and reduces the neutron capture time (Fig. $\ref{fig:fig8b}$). Since the neutron capture time determines the size of the prompt-delayed time window, early capture of neutrons shortens this time interval and hence reduces accidental backgrounds. However, raising the efficiency and reducing the capture time in this way can be efficient to some extent since it becomes more and more difficult to change these values by increasing the Gd-layer thickness. Therefore, a Gd$_2$O$_3$ layer that is thick enough to ensure high neutron capture is selected. We choose 50 $\mu$m thick Gd$_2$O$_3$ layer (as in Panda design) for the wrapping and use this value for the subsequent simulations.

\begin{figure}[!htb]
    \centering
    \subfigure[Mean of the distribution of neutron capture times. ]
    {
        \includegraphics[width=0.47\linewidth]{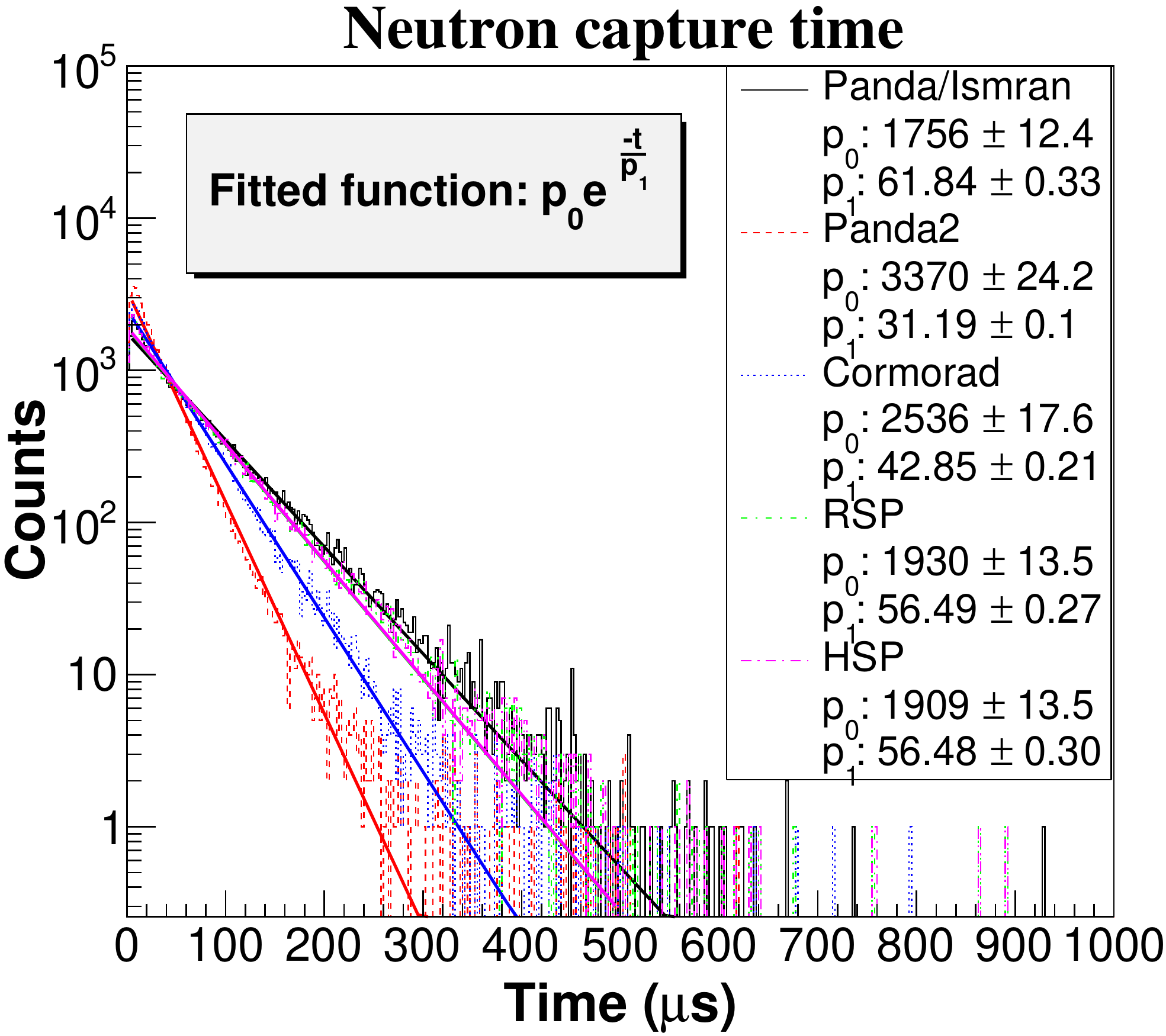}
        \label{fig:fig9a}
    }
    \subfigure[ Efficiency of neutron capture as a function of capture time.]
    {
       \includegraphics[width=0.47\linewidth]{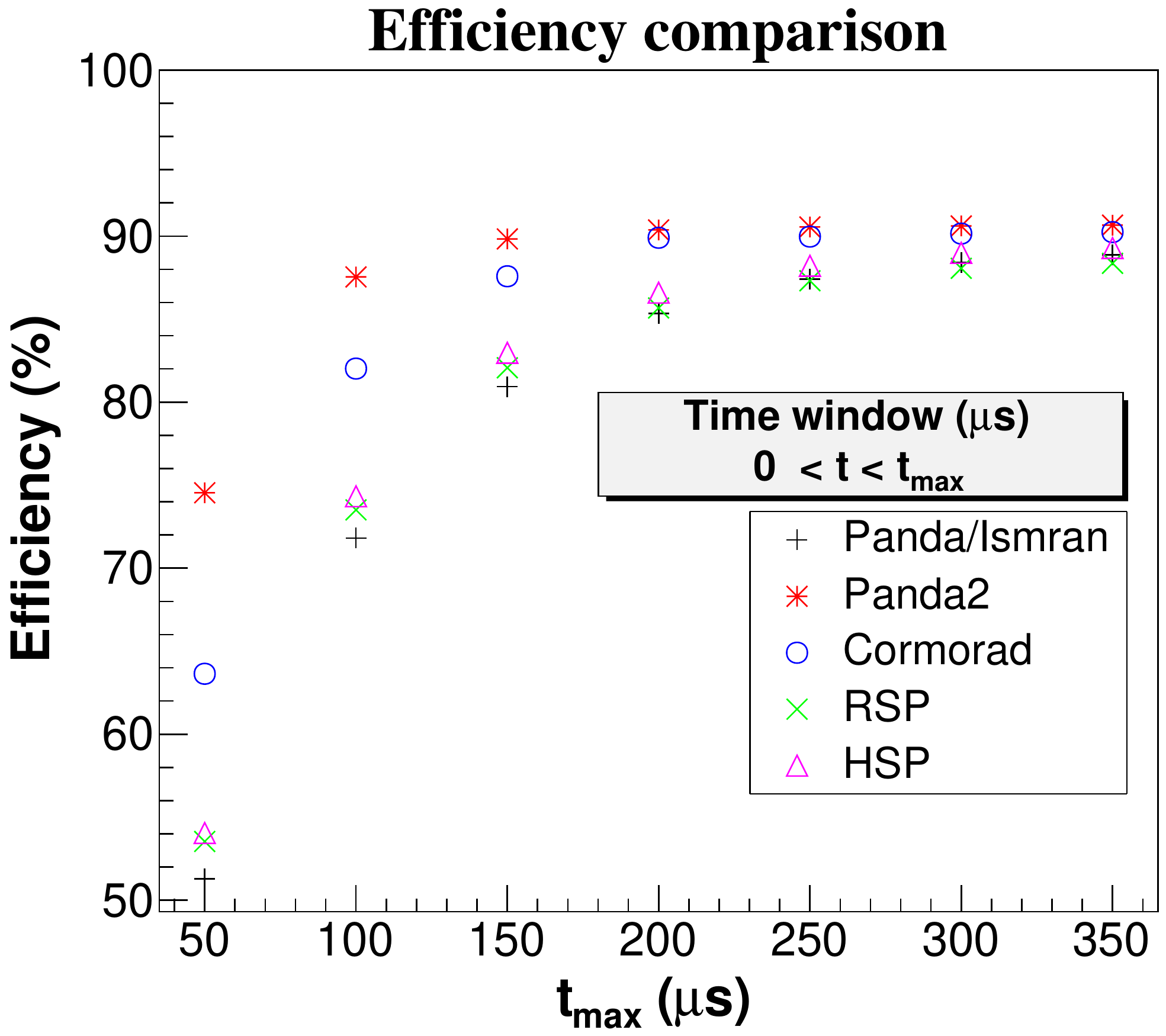}
        \label{fig:fig9b}
    }
    \caption{Comparison of different detector designs. The designs differ from each other in terms of the size and shape of the modules used. }
    \label{fig:fig9}
\end{figure}

Fig. $\ref{fig:fig9a}$ shows the distribution of neutron capture time for each detector design. All the distributions are fitted to a two-parameter exponential function to obtain the mean neutron capture time. The obtained fit parameter $p_1$ gives the mean neutron capture time. The mean capture time and the neutron capture efficiency (for both gadolinium and hydrogen) of each detector design are reported in table $\ref{table:table3}$. Furthermore, the variation of neutron capture efficiency depending on the correlated time window is shown in Fig. $\ref{fig:fig9b}$. For a quantitative comparison of the neutron capture time, the resulting size of the time window to collect $85\%$ of the neutrons is calculated for each detector and the results are presented in table $\ref{table:table3}$.

\begin{figure}[!htb]
    \centering
    \subfigure[Plastic antineutrino detector array (Panda).]
    {
        \includegraphics[width=0.31\linewidth]{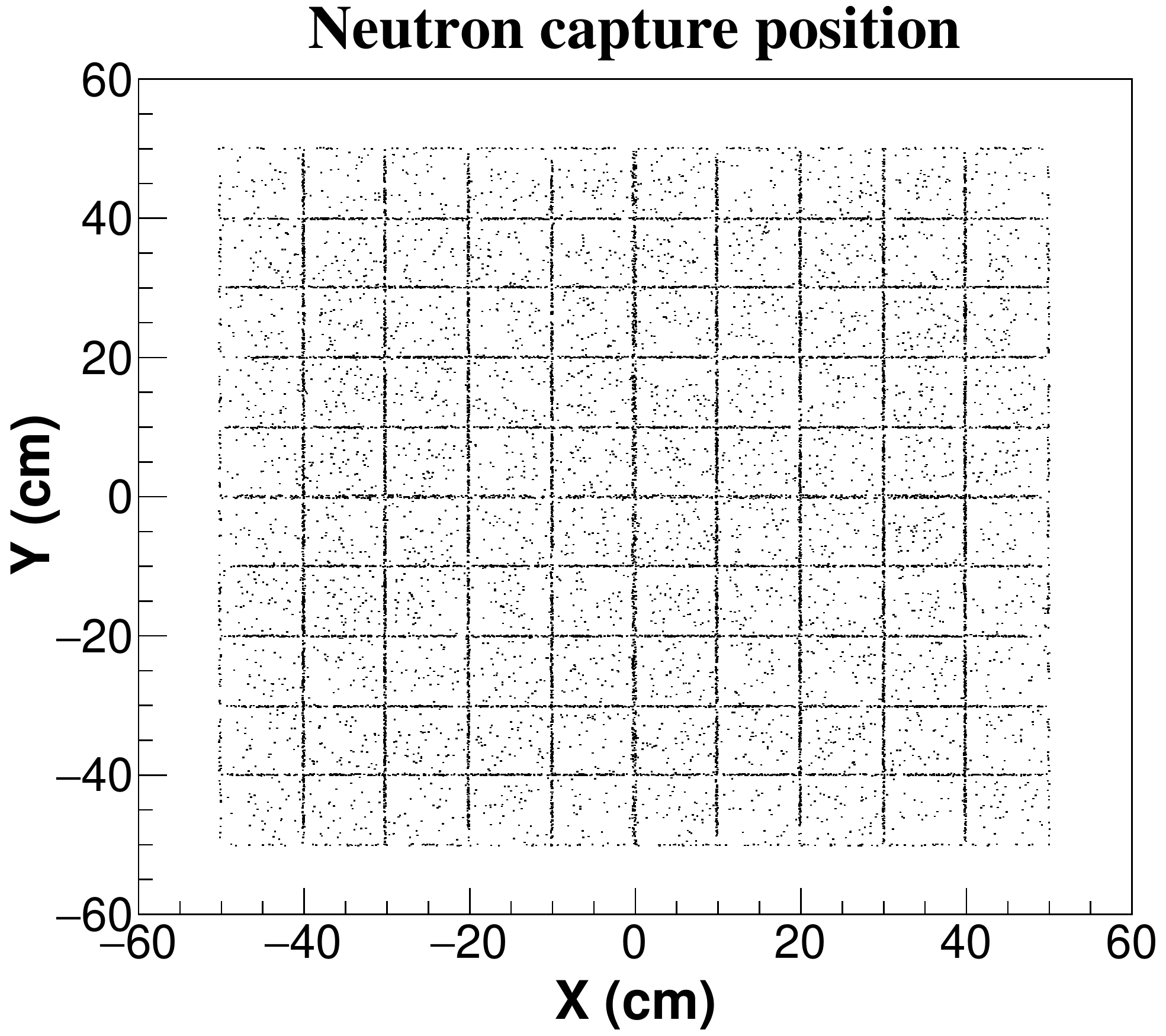}
        \label{fig:fig10a}
    }
     \subfigure[Rectangular shaped packing (RSP).]
    {
       \includegraphics[width=0.31\linewidth]{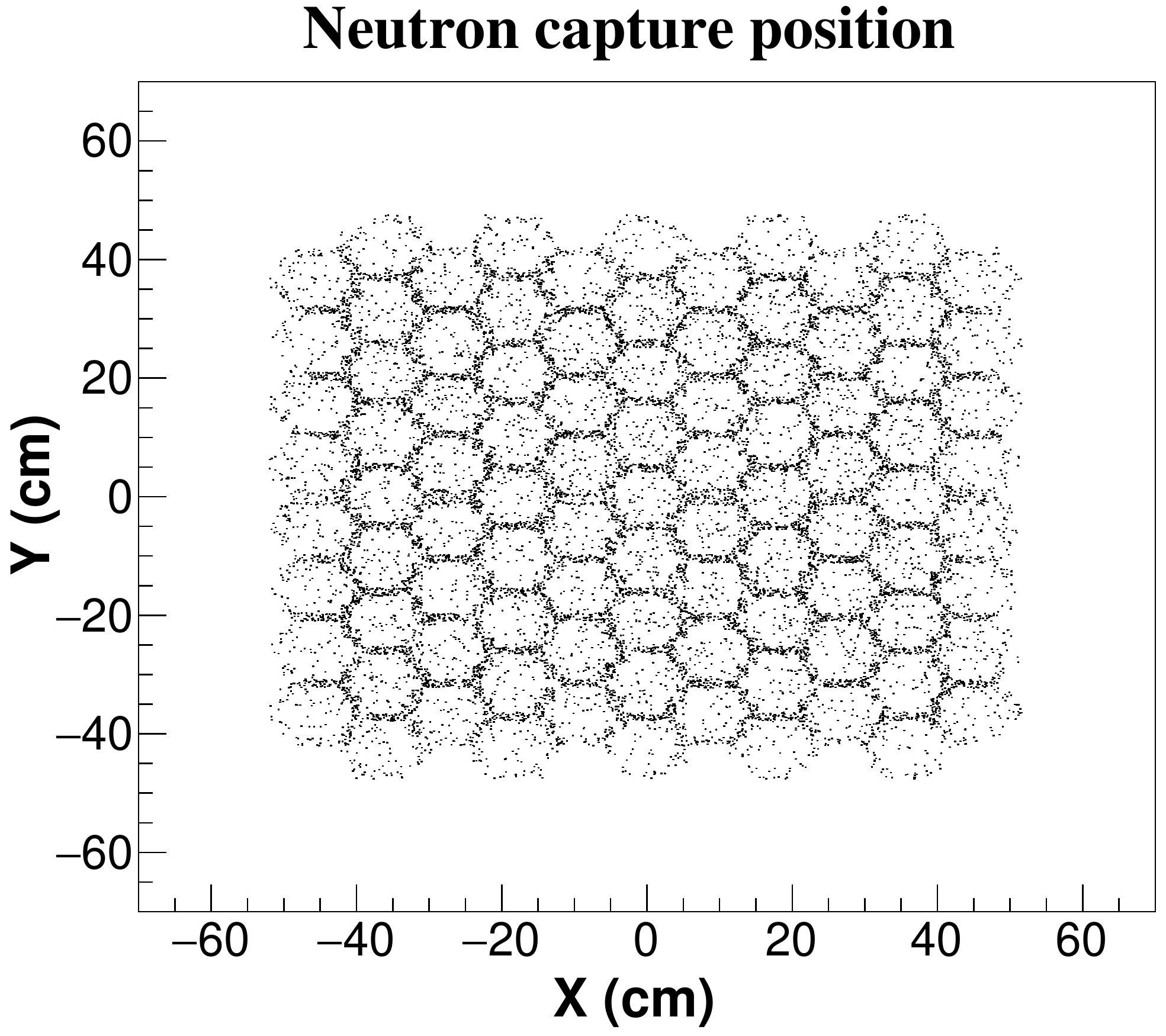}
        \label{fig:fig10b}
    }
    \subfigure[Hexagonal shaped packing (HSP).]
    {
       \includegraphics[width=0.31\linewidth]{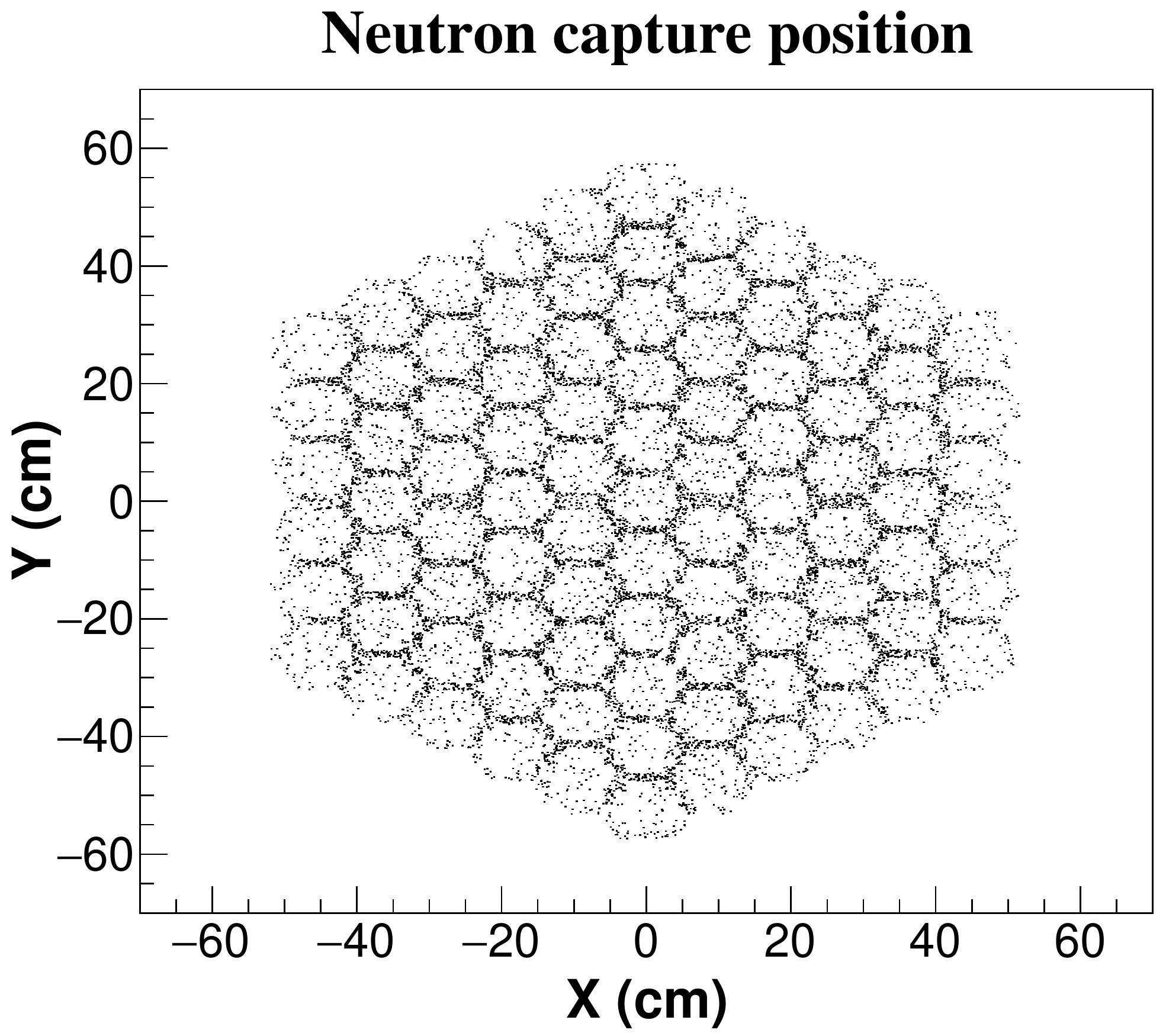}
        \label{fig:fig10c}
    }
    \caption{ Cross-sectional view of the neutron capture profile of three different detector configurations. }
    \label{fig:fig10}
      
\end{figure}

Fig. $\ref{fig:fig10}$ shows the profile of neutron capture position for Panda (Fig. $\ref{fig:fig10a}$), RSP (Fig. $\ref{fig:fig10b}$) and HSP (Fig. $\ref{fig:fig10c}$) designs. Since the surface of the modules is wrapped with gadolinium-oxide and the thermal capture cross-section of gadolinium is very high compared to that of hydrogen, neutrons are mostly captured near the edges of the plastic bars. The capture profile of the Cormorad and Panda2 are not shown since they have a similar pattern with Panda detector. 

Neutron capture efficiency and time are also depended on how the Gd is distributed over the active volume of the detector. Instead of wrapping plastic bars with Gd, dissolving Gd homogeneously in plastic scintillator (e.g., 0.1 wt$\%$) reduces mean capture time (from 62 to 30 $\mu$s for Panda design) and raises efficiency \cite{Oguri1}. However, this method comes with a drawback. It increases the quenching effects (reduced light output) of the scintillator hence leads to the energy resolution of the detector to decrease.

\subsubsection{Antineutrino detection efficiency}

In order to compare different design alternatives in terms of antineutrino detection efficiency, we apply the same simple energy and time cuts for each type of the detector. Since the aim is to find the most efficient design, these simple selection cuts are sufficient to determine the most suitable design for $\overline{\nu}$ detection. In a real case, further cuts are definitely necessary to discriminate antineutrino events from antineutrino-like background events which are very high compared to $\overline{\nu}$ events, especially at above ground condition.

\begin{figure}[!htb]
    \centering
    {
        \includegraphics[width=0.47\linewidth]{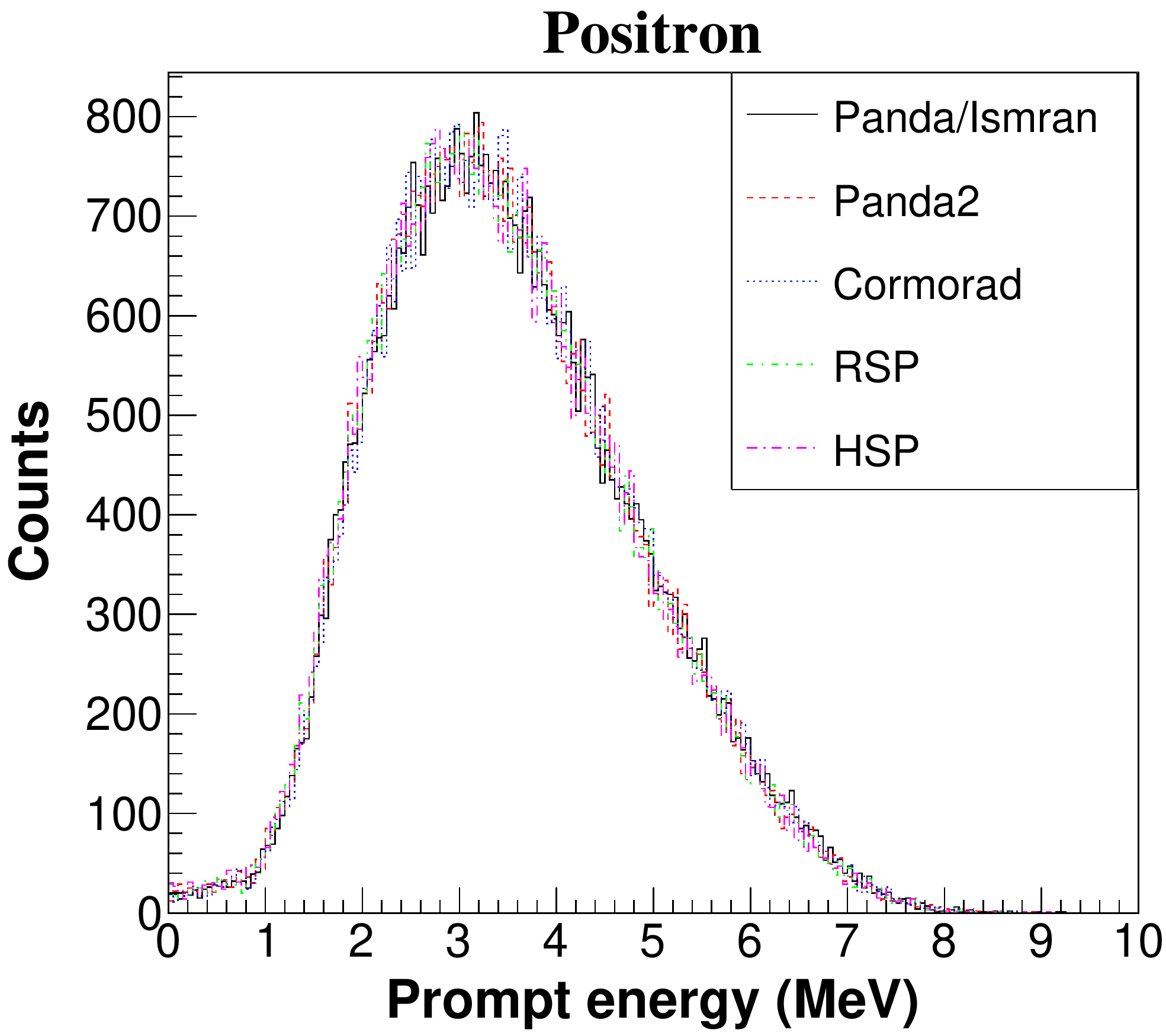}
        \label{fig:fig11a}
    }
    {
       \includegraphics[width=0.47\linewidth]{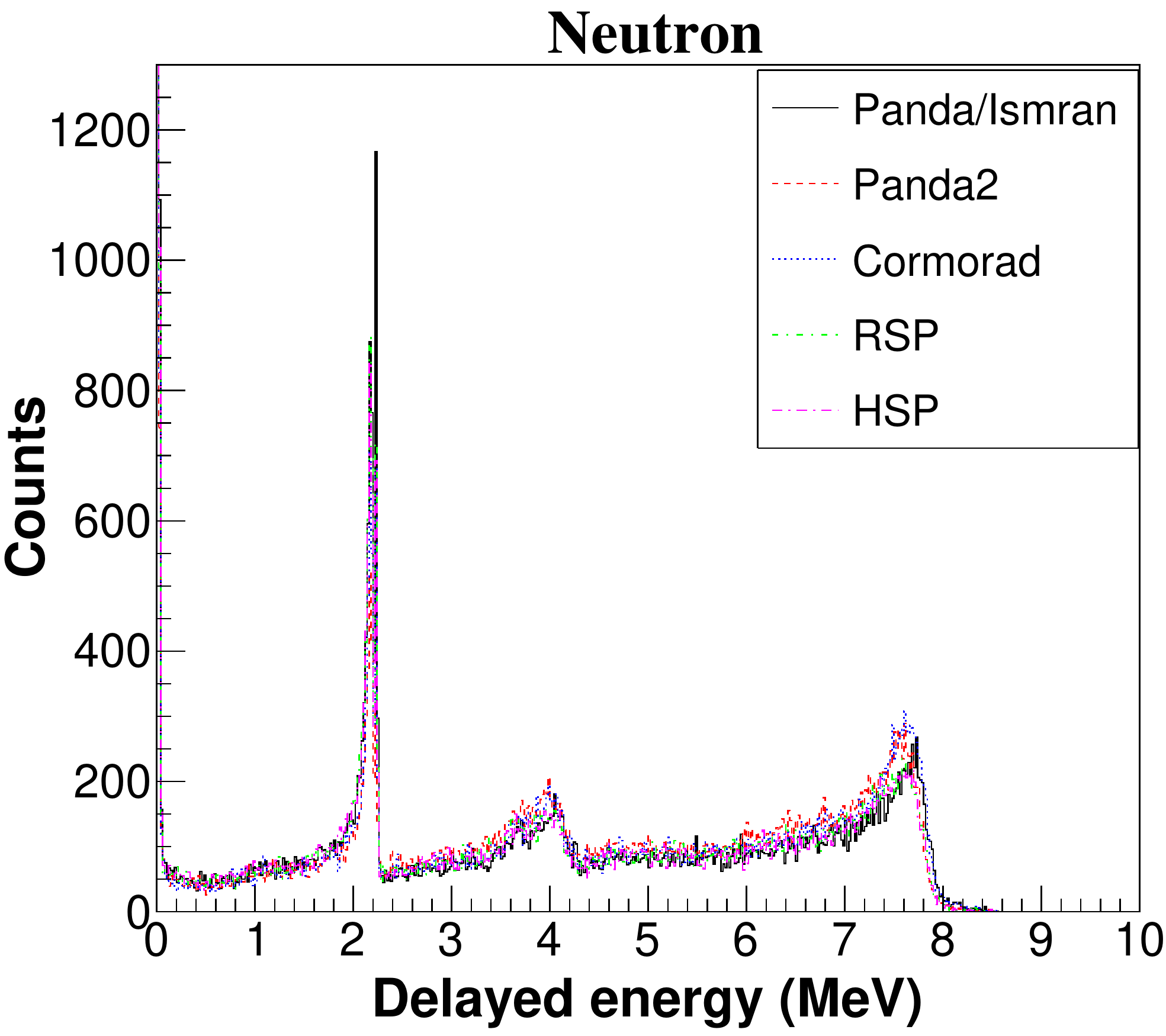}
        \label{fig:fig11b}
    }
    \caption{ Total deposited energy arises from the positron and neutron. The little relative shifts in the delayed energy spectra result from the Gd content of the detectors.  }
    \label{fig:fig11}
\end{figure}

\begin{figure}[!htb]
    \centering
    {
        \includegraphics[width=0.47\linewidth]{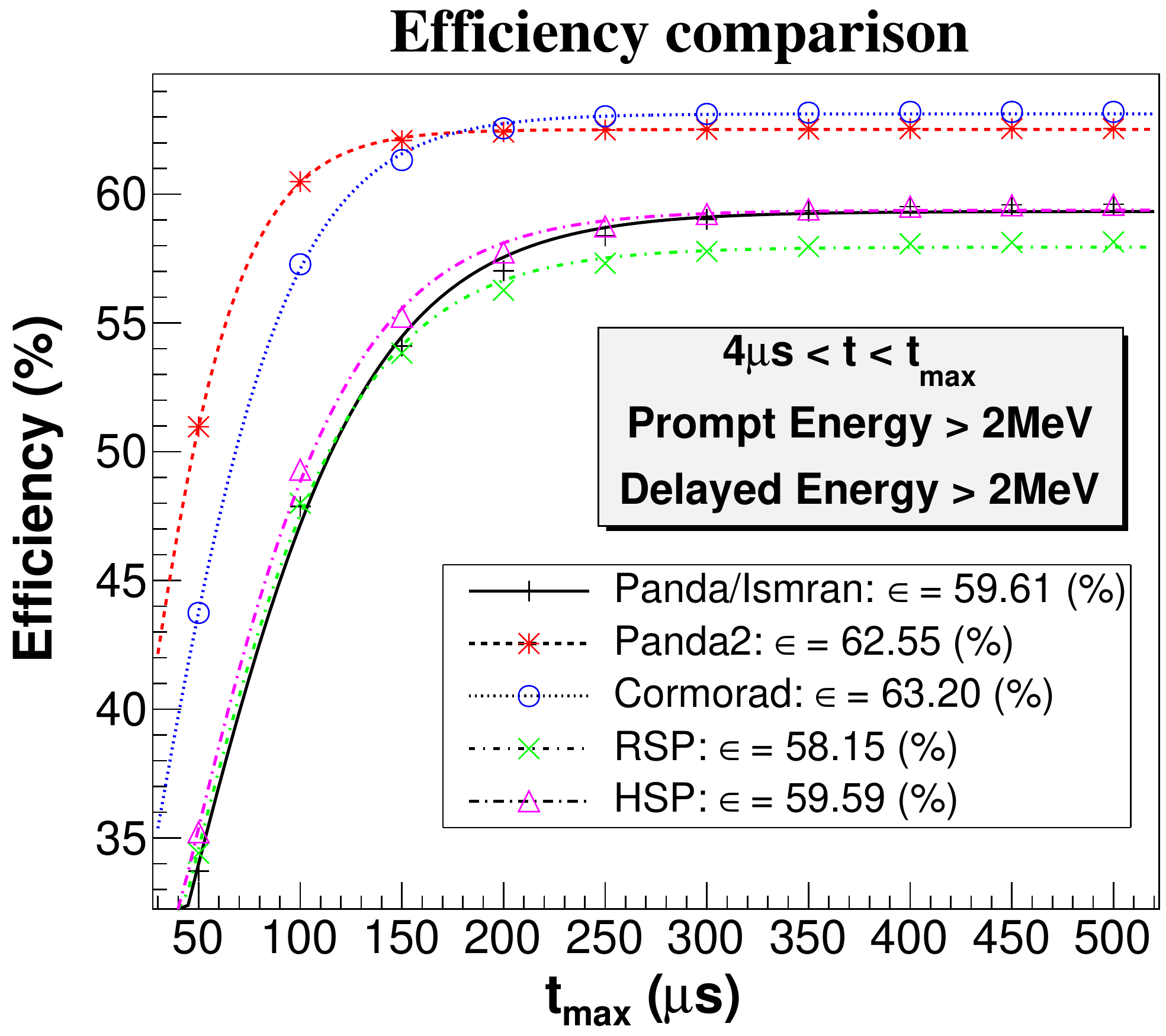}
    }
    {
       \includegraphics[width=0.47\linewidth]{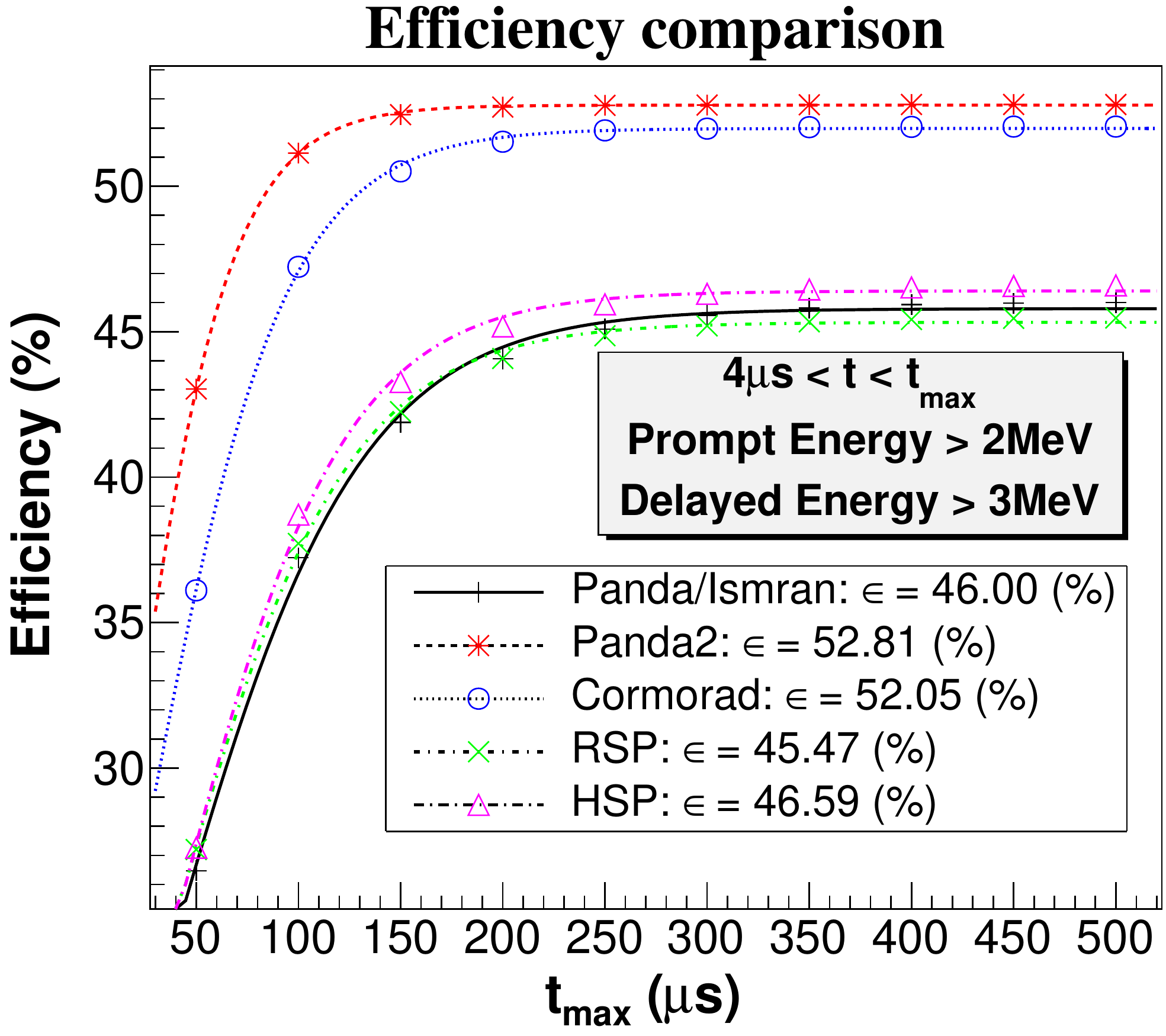}
    }
    \\
    {
       \includegraphics[width=0.47\linewidth]{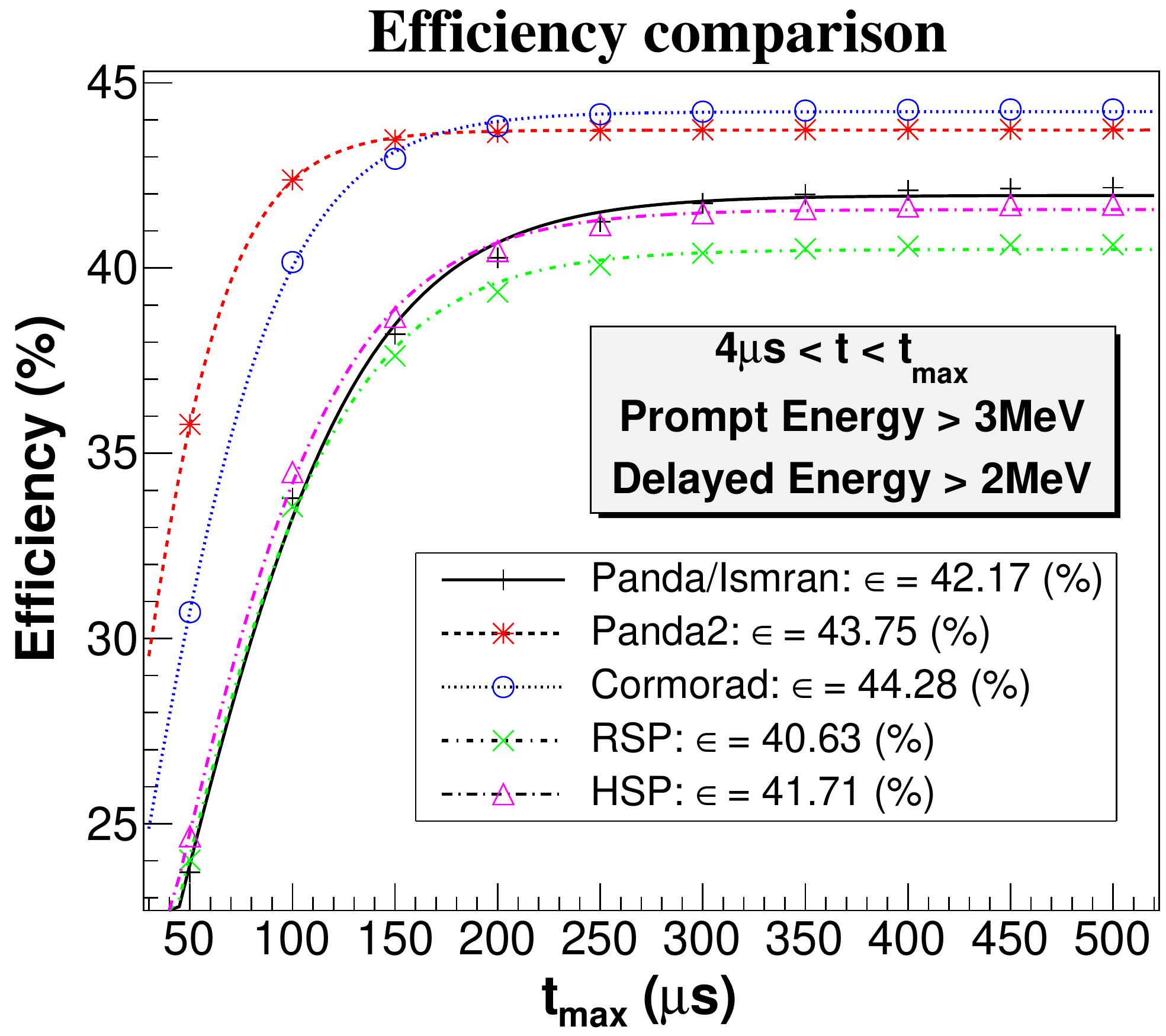}
    }
    {
       \includegraphics[width=0.47\linewidth]{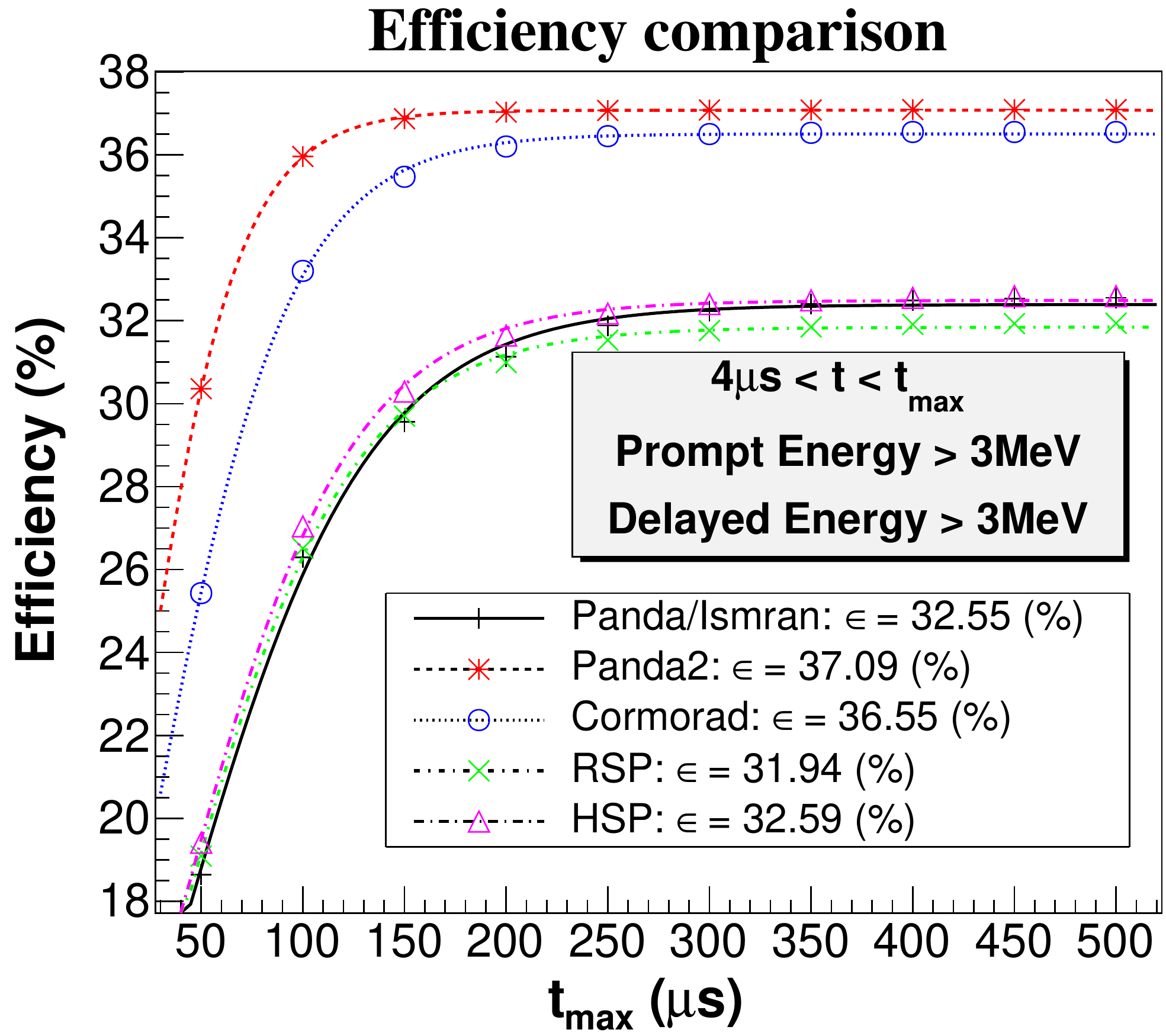}
    }
  
  \caption{ Variation of $\overline{\nu}$ detection efficiency as a function of neutron capture time window.  }
    \label{fig:fig12}
\end{figure}

Fig. $\ref{fig:fig12}$ compares the response of each detector type when simple prompt-delayed energy and time cuts are applied to the deposited energy (Fig. $\ref{fig:fig11}$) and neuton capture data (Fig. $\ref{fig:fig9a}$). Panda2 and Cormorad design show similar behavior and they exhibit better performance than Panda/Ismran and HSP/RSP design since they achieve to reach a higher efficiency value in a shorter time. HSP/RSP design reach nearly the same efficiency value with Panda/Ismran design but time to reach the saturation is slightly shorter. All the data points in each figure are fitted to a sigmoid function to find the point where the efficiency saturation begins. By doing this, we obtain the smallest correlation time window without sacrificing efficiency. As an example, the obtained approximate values from Fig. $\ref{fig:fig12}$ (bottom right) are shown in table $\ref{table:table3}$. We also report the efficiency value obtained in two time intervals in order to have a clearer comparison of the different detector performances.

Another point to pay attention to is the effect of Gd$_2$O$_3$ layer thickness on the detection efficiency. Although increasing the thickness initially increases the efficiency, after a certain thickness value, increasing the thickness reduces the efficiency since the possibility of few MeV positrons (which has a few cm ranges) hitting the wrapping material and thus losing the energy information increases. More clearly, as the thickness increases, prompt efficiency always decreases while the delayed efficiency always increases. If the thickness is increased more than necessary, the prompt contribution dominates, leading to a decrease in the detection efficiency. Therefore, it is important to determine the thickness value that gives the maximum efficiency. We find the optimum thickness of Gd$_2$O$_3$ layer to be around 50-60 $\mu$m for all designs.

\subsubsection{Detected antineutrino event rate}

\begin{figure}[!htb]
\centering
 \includegraphics[width=0.47\linewidth]{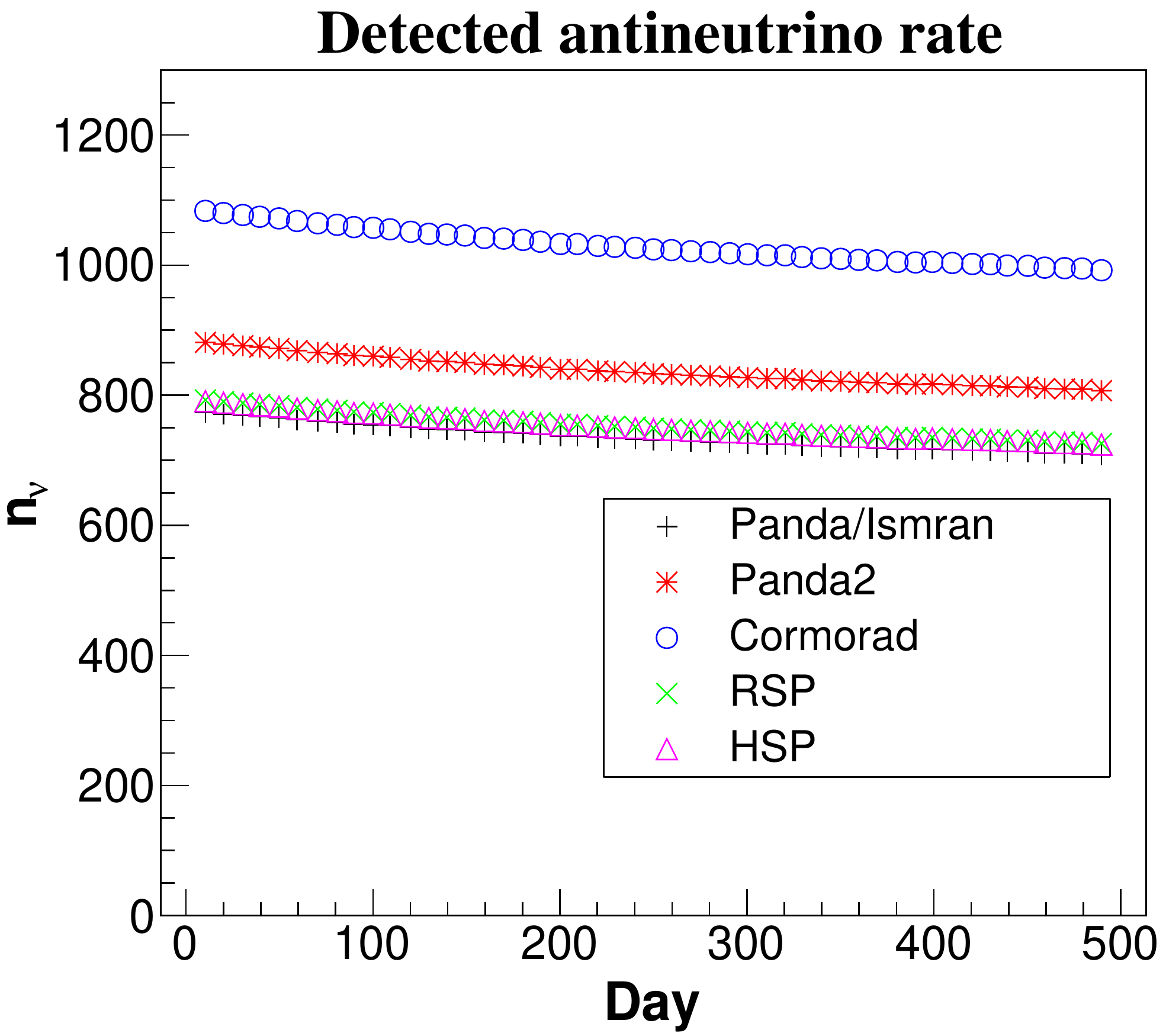}
\caption{Evolution of the detected antineutrinos over the fuel cycle. Since the volume of the Cormorad contains more proton target than other designs, the number of antineutrinos expected to be detected is higher. }
\label{fig:fig13}
\end{figure}

To see the relation between the antineutrino count rate and the core composition of the reactor, the number of detected antineutrinos is calculated for each type of the detector by using equation $\ref{eq:2}$. The detector is considered to be placed 30 m away from a 1 GW power reactor for the calculation. In the calculation, the change of fission fraction of four isotopes with time is taken from Fig $\ref{fig:fig2a}$. The parameters describing the detectors and the cross section per fission values of each of the four isotopes are taken from table $\ref{table:table1}$ and table $\ref{table:table3}$ respectively. Fig. $\ref{fig:fig13}$ shows the change of the detected antineutrinos for each type of the detector over the fuel cycle.

\subsection{Optical photon }
\label{sec:opticPhysics}

\begin{figure}[!htb]
    \centering
    
    \subfigure[ The peak at energy 1511 and 2022 keV correspond to the energy deposition of 1 MeV positron plus one or two annihilation gamma respectively. ]
    {
       \includegraphics[width=0.45\linewidth]{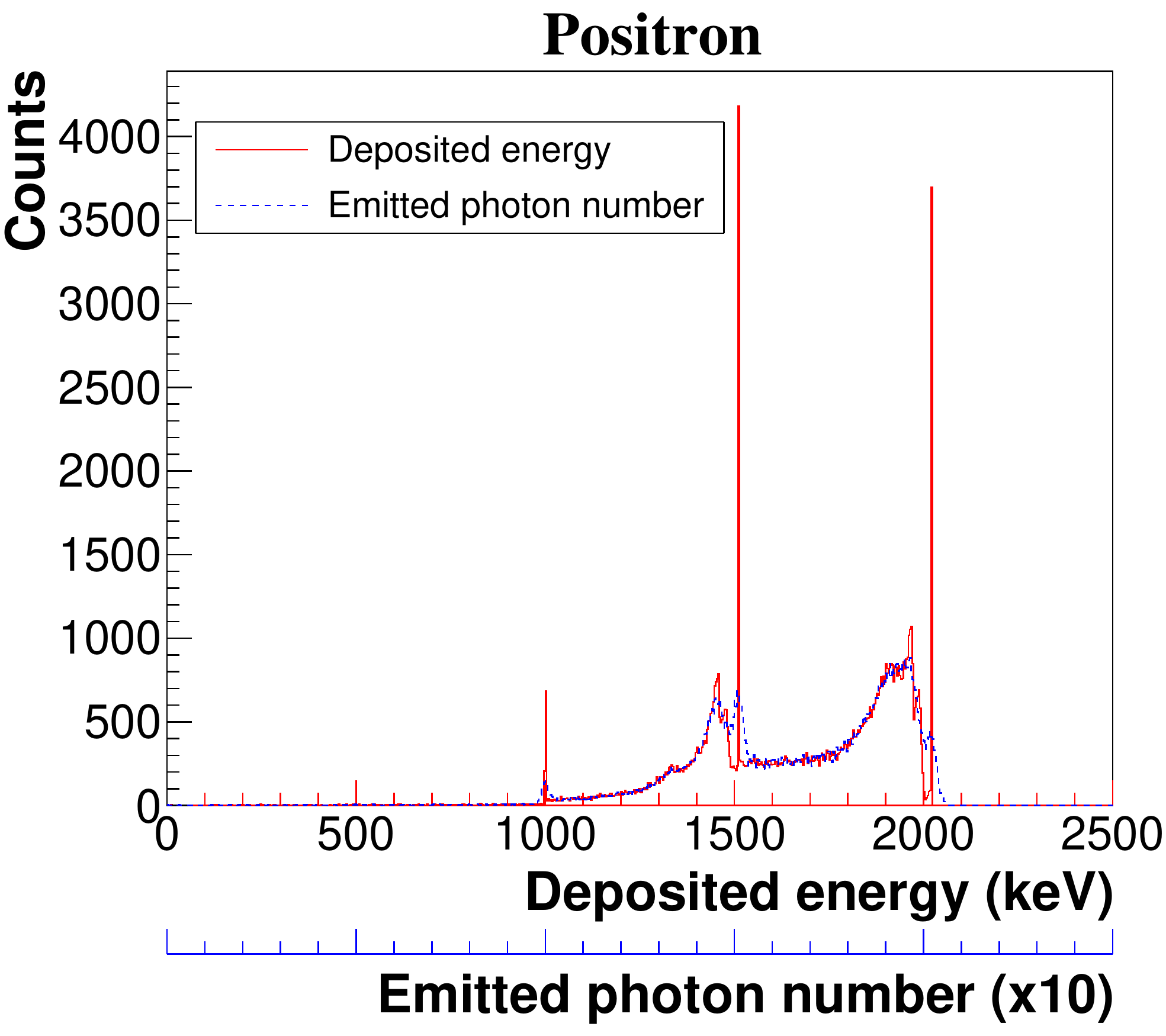}
        \label{fig:fig14a}
    }      
    \quad
     \subfigure[Visual representation of positron simulation in HSP detector. The red lines represent annihilation gamma and the blue lines represent optical photons.]
    {
       \includegraphics[width=0.45\linewidth]{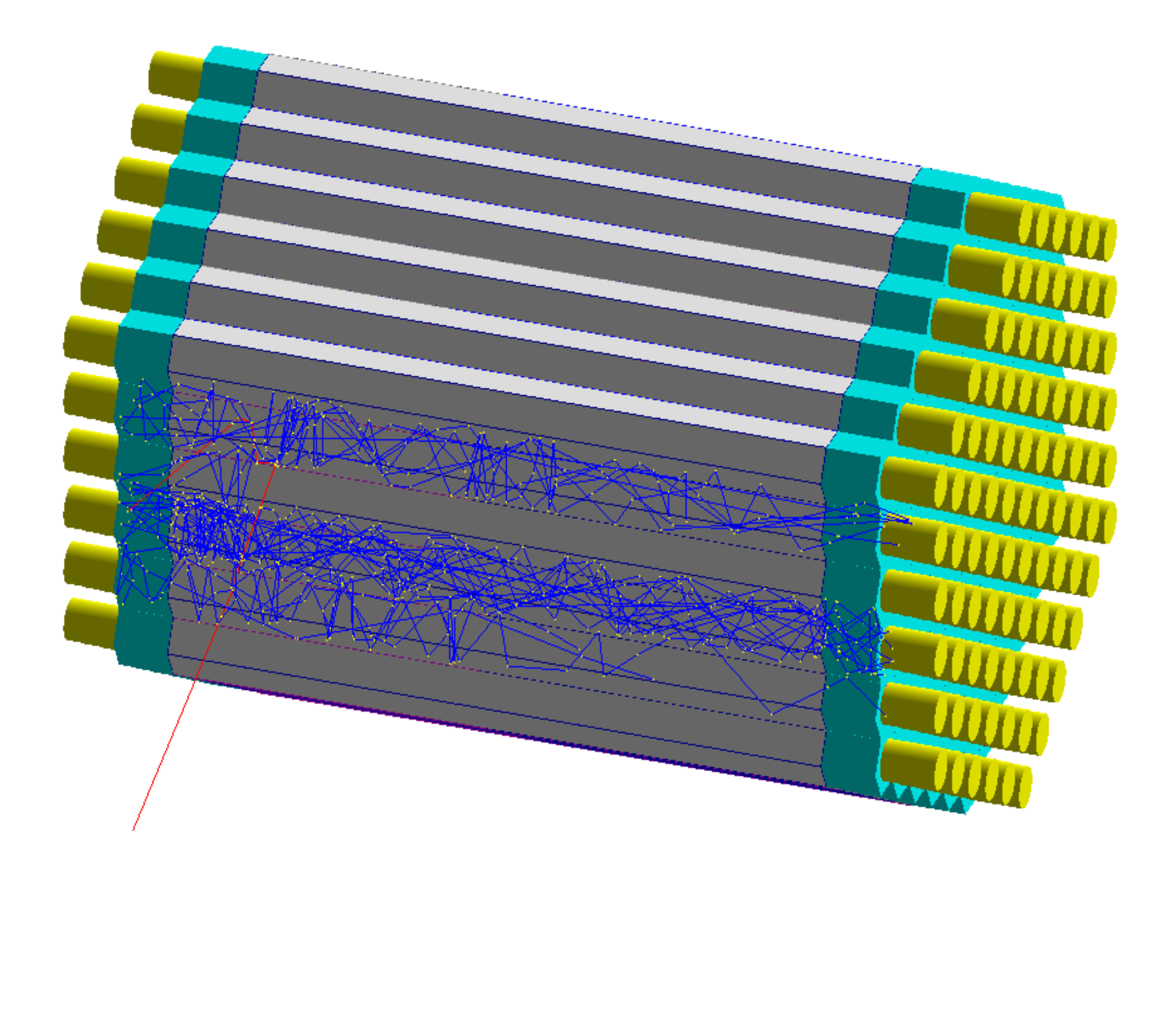}
        \label{fig:fig14b}
    }
    
    \caption{ Simulation of 1 MeV positron energy deposition. When positron deposits E amount of energy in the scintillator (EJ-200), N scintillation photons are emitted with an expectation value of $\mu_{N} = E $ (keV) x  (Scintillation yield = 10 photons/KeV) and a standard deviation of $\sqrt{\mu_{N}}$.  }
    \label{fig:fig14}
      
\end{figure}

In order to compare the performance of the detectors in terms of energy resolution, we simulate energy deposition of mono-energetic positrons for each detector type. A positron is generated in a random position within the detector active volume and then fired in any direction. Following the energy deposition of the positron in the scintillator, optical photons are emitted isotropically and tracked throughout the detector volume until they disappear with one of these processes: self-absorption in the scintillator medium, losses at the surface boundary and absorbed or detected by the PMTs. Fig. $\ref{fig:fig14a}$ shows the energy deposition of 1 MeV positron (solid line) and the emitted photon number (dashed line) versus the deposited energy. A schematic illustration of the simulation process is also shown in Fig. $\ref{fig:fig14b}$.

\begin{figure}[!htb]
    \centering
  \subfigure[Module type 1,2,3 and 4 correspond to modules of Panda, Panda2, Cormorad, and HSP/RSP respectively.]
    {
       \includegraphics[width=0.45\linewidth]{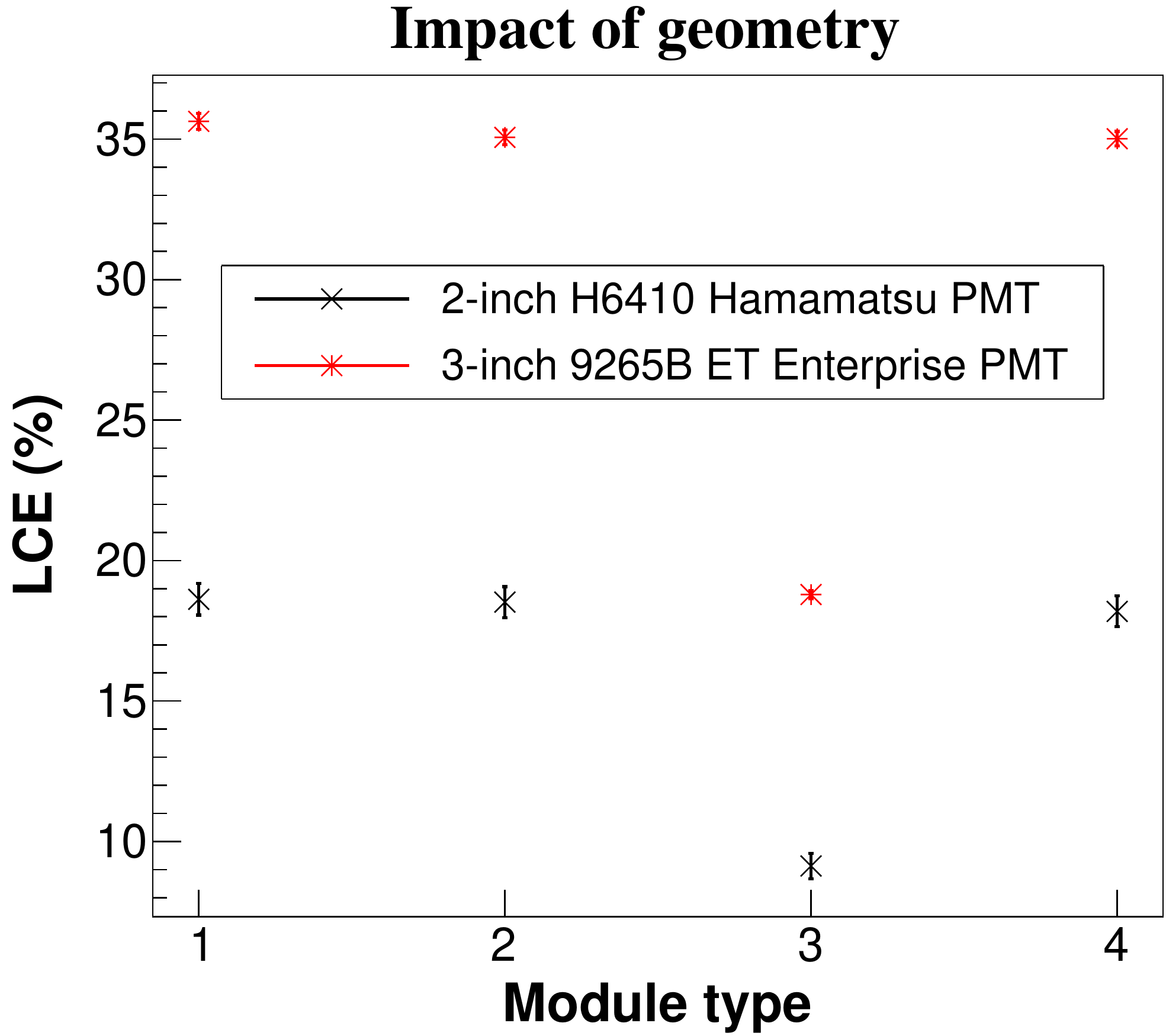}
        \label{fig:fig15a}
    }
    \quad
    \subfigure[1 MeV photoelectron spectrum for each detector type. ]
    {
       \includegraphics[width=0.45\linewidth]{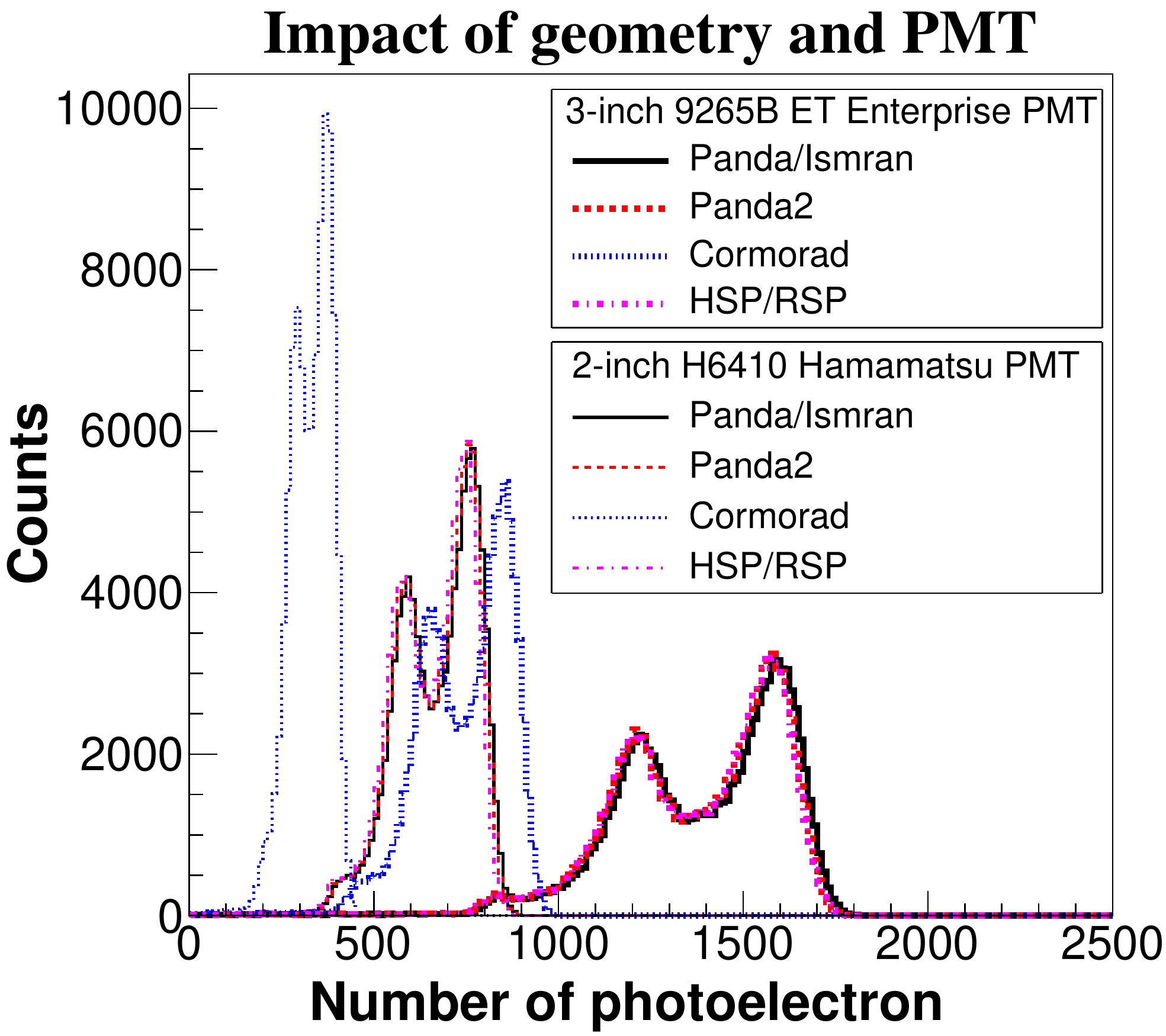}
        \label{fig:fig15b}
    }    
    
    \caption{ The impact of module shape and size on the light collection and detection efficiency of the detectors in the case of using two different PMT.  }
    \label{fig:fig15}
      
\end{figure}

Since the best energy resolution is obtained by collecting the maximum number of photons, we first investigate the light collection efficiency (LCE) of each detector. Fig. $\ref{fig:fig15a}$ shows the average light collection efficiency (i.e., ratio of optical photons arriving at the photocathode to the total number of photons generated in the scintillator) of the detectors in the case of using two different PMTs. The most important result in Fig. $\ref{fig:fig15a}$ is that the HSP/RSP design collects an equal amount of light with the Panda design ($18\%$ for 2-inch and $35\%$ for 3-inch) although it is equipped with $9\%$ fewer PMTs (182 vs 200). On the other hand, Cormorad collects only half of the light collected by other detectors since it is instrumented by 98 PMTs. What is important here is to minimize the number of PMTs used for a given detector volume while achieving the same amount of light collection. In this respect, the HSP design exhibits the best performance. Furthermore, a significant increase in LCE arises from the size of the photocathode. Increasing the photocathode radius from 2.3 cm (2-inch H6410) to 3.5 cm (3-inch 9265B) increases the LCE by about twice. The main losses in the light collection process are due to the scintillation photons absorbed in the scintillator medium. If we take into account the quantum efficiency effect of the PMT's photocathode, the number of detected photons is obtained. Fig. $\ref{fig:fig15b}$ shows the photoelectron number distribution of each detector for 1 MeV positron energy deposition in the case of using two different PMTs. The results are also shown in table $\ref{table:table3}$.

\begin{figure}[!htb]
    \centering
    {
       \includegraphics[width=0.47\linewidth]{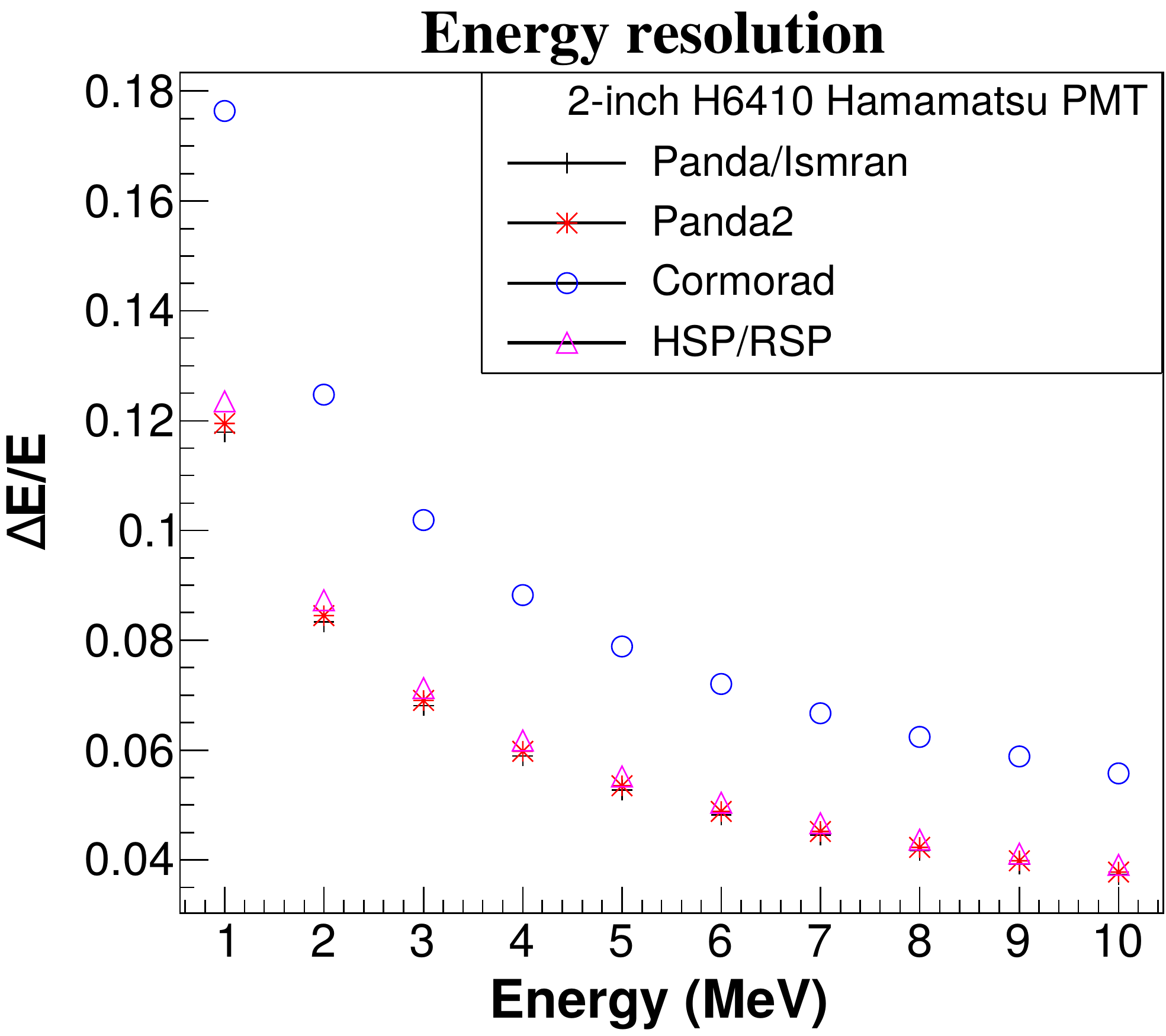}
    }
    {
       \includegraphics[width=0.47\linewidth]{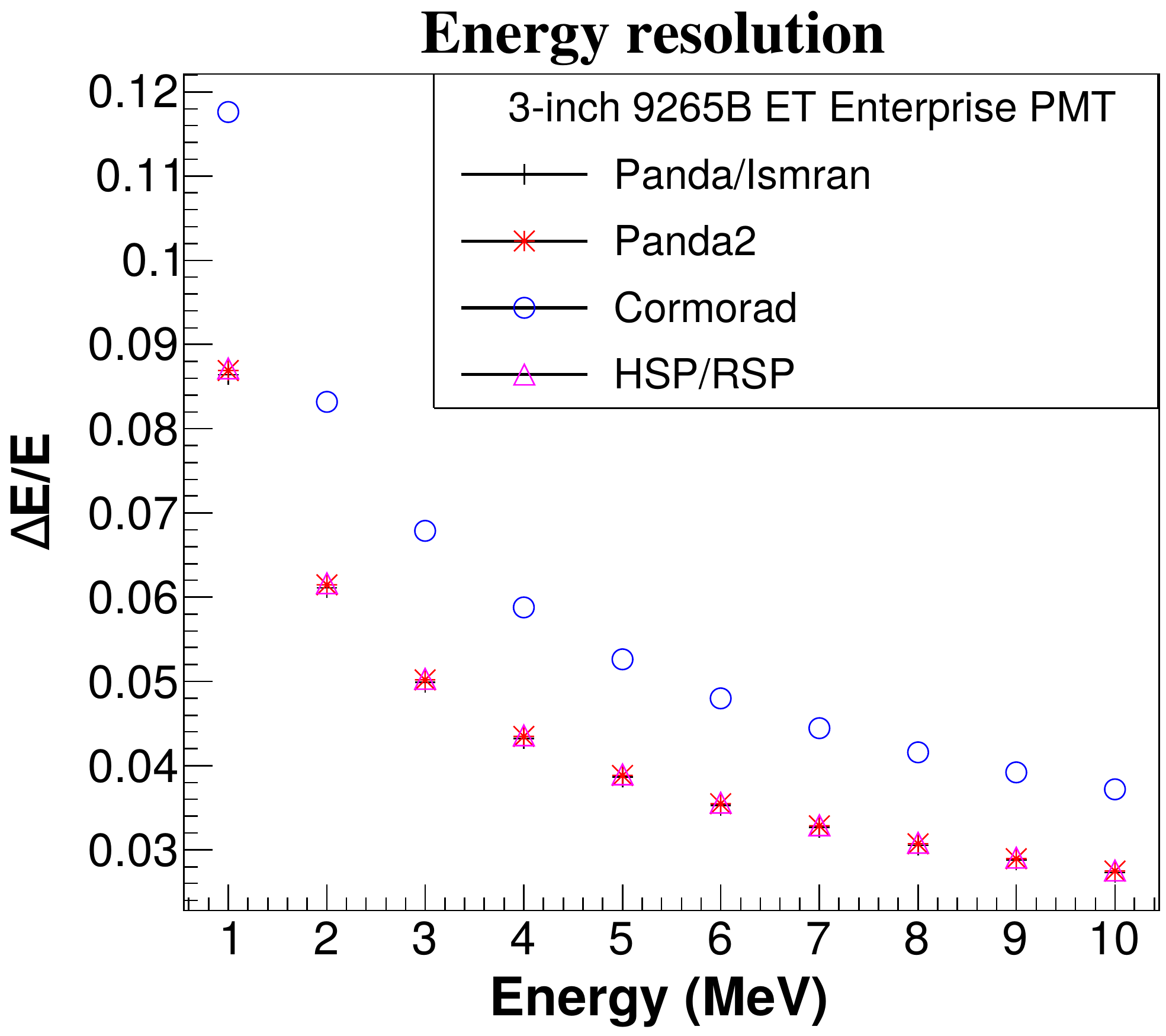}
    }    
    
    \caption{ The impact of module shape and size on the energy resolution of the detectors in the case of using two different PMT. }
    \label{fig:fig16}
      
\end{figure}

Energy resolution of each detector is calculated using the photoelectron number distribution, which includes the fluctuation of scintillation process, light collection process and detection process. The impact of photoelectron collection and multiplication on energy resolution is not considered in the simulation. Fig. $\ref{fig:fig16}$ compares the resolutions of the detectors in the case of using two different PMTs.

\clearpage

\begin{longtable}[!htb] {|p{3.2cm} |p{2.8cm} | p{1.8cm} | p{2.0cm} | p{1.8cm} | p{1.8cm} |}

\caption{ Summary of the simulation results. The abbreviations LCE and LDE stand for light collection and detection efficiency, respectively. } 
\label{table:table3} \\
   
\endfirsthead
\caption* {\textbf{Table \ref{table:table3} (continued)} }\\
\endhead    
 

\hline
\multirow{2}{*}{ \shortstack[l]{ \textbf{Quantity} }  } & \multicolumn{3}{|c|}{ \textbf{Existing Design} } & \multicolumn{2}{|c|}{ \textbf{Proposed Design} } \\
\cline{2-6}

                   & \textbf{Panda/Ismran} & \textbf{Panda2}  & \textbf{Cormorad}  &  \textbf{HSP}  &  \textbf{RSP}   \\
 \hline
 
 \multicolumn{6}{|c|}{ \textbf{Description of detectors} } \\
 \hline
 
 x (cm) &  10 & 5 & 7 & 6 & 6  \\
 \hline
 y (cm) &  10 & 5 & 7 & - & -  \\
 \hline
 z (cm)          &  100  & 100  & 130  & 120  & 120  \\
 \hline
 $N_m$           &  100  & 100  & 49   & 91   & 93  \\
 \hline
 $N_{Pmt}$       &  200  & 200  & 98   & 182  & 186  \\
 \hline
 Mass (kg)        &  1023 & 1023 & 1277 & 1045 & 1068 \\
 \hline
 Volume ($m^3$)  &  1.00 & 1.00 & 1.25 & 1.02 & 1.04 \\
 \hline
 $N_p$ $(x10^{28})$ &  5.17 & 5.17 & 6.45 & 5.28 & 5.40 \\
 \hline
 \raggedright Total Gd concentration $\%$(w/w) &  0.19 & 0.38 & 0.27 & 0.18 & 0.18 \\
 \hline
 
  \multicolumn{6}{|c|}{ \textbf{Simulation results} } \\
  \hline
\multicolumn{6}{|c|}{ \textbf{Relevant parameters of neutron} } \\ 
 \hline
 Hydrogen ($\%$)   & 22.05  &  11.02   &  14.89   & 17.88    &  17.68   \\
 \hline
 Gadolinium ($\%$) & 66.93  &  79.54   &  75.20   & 70.98    &  70.30   \\
 \hline
 Escape ($\%$)     & 10.61  &  9.28    &  9.66    &  10.85   &  11.67   \\
 \hline
  
\raggedright Mean neutron capture time ($\mu$s) & 61.84 & 31.19  & 42.85 &  56.48 &  56.49   \\
\hline
\raggedright  Required size of the time window to collect $85\%$ of the IBD neutrons ($\mu$s)  & 195  &  81   &  120   & 175    &  186   \\
 \hline
\multicolumn{6}{|c|}{ \textbf{Antineutrino detection efficiency} } \\ 
\multicolumn{6}{|c|}{ \textbf{Prompt energy $>$ 3 MeV and Delayed energy $>$ 3MeV} } \\
 \hline
 
 \raggedright Minimum correlated time ($\mu$s)   & 260  &  150   &  200   & 240      &  240   \\
 
 \hline
\raggedright Maximum efficiency ($\%$)          & 31.99  &  36.87    &  36.20    &  32.10   &  31.46   \\
 \hline
\raggedright Time window ($\mu$s) & \multicolumn{5}{|c|}{ \textbf{Efficiency ($\%$)  } } \\
 \hline 
 
\raggedright  4 $<$ t $<$  50   & 18.64  &  30.36    &  25.43    &  19.41   &  19.09 \\
 \hline 
 
\raggedright  50 $<$ t $<$ 100   & 7.65  &  5.61    &  7.77    &  7.64   &  7.41  \\
 \hline  
 
 \raggedright 100 $<$ t $<$ 150   & 3.27  &  0.91    &  2.26    &  3.25   &  3.19  \\
 \hline
 
 \raggedright 150 $<$ t $<$ 200   & 1.57  &  0.16    &  0.73    &  1.35   &  1.29  \\
 \hline
 
 \raggedright 200 $<$ t $<$ 250   & 0.36  &  0.01    &  0.05    &  0.24   &  0.23  \\
 \hline
 
  \raggedright 250 $<$ t $<$ 500   & 0.32  &  0.01    &  0.05    &  0.19   &  0.18  \\
 \hline
 
\multicolumn{6}{|c|}{ \textbf{Optical photon parameters} } \\  
 \hline
 
\multicolumn{6}{|c|}{ \textbf{2-inch H6410 Hamamatsu PMT}\cite{Hamamatsu} } \\
 \hline
  LCE ($\%$)      & 18.62     &  18.52    &  9.13    &  18.19   &  18.18   \\
 \hline
  LDE ($\%$)   & 4.00  &  3.97   &  1.97   & 3.91    &  3.90   \\
 \hline
 Energy resolution ( $ \%/\sqrt{E (MeV)}$ )   & 11.79  &  11.95   &  17.64   & 12.35 &  12.35   \\
 \hline
  \multicolumn{6}{|c|}{ \textbf{3-inch 9265B ET Enterprise PMT}\cite{Et} } \\
 \hline
  LCE ($\%$)      & 35.63     &  35.06    &  18.79    &  35.01   &  34.97   \\
 \hline
  LDE ($\%$)   & 8.30  &  8.21   &  4.45   & 8.19    &  8.16   \\
 \hline 
 Energy resolution ( $ \%/\sqrt{E (MeV)}$ )   & 8.64  &  8.69   &  11.76   & 8.72    &  8.72   \\
 \hline

 
\end{longtable}

\clearpage

\section{Conclusion}

In this study, the most efficient plastic antineutrino detector design is investigated for near-field reactor monitoring application. The detector designs, which are used in a few ambitious experiments, are compared with our new geometrically enhanced design in terms of antineutrino detection and energy resolution efficiency. In addition, the influence of important design parameters on the performance of the detectors is explained comparatively.

Firstly, the impact of Gd$_2$O$_3$ layer thickness on the neutron capture efficiency and the neutron capture time are investigated. The results show that increasing the thickness raises the capture efficiency and reduces the capture time. For the antineutrino detection efficiency, which includes both the prompt and delayed efficiency as well as the correlated time, it is found that there is an optimum thickness value at which the maximum efficiency is obtained. Increasing the thickness up to this value significantly improves the detection efficiency. If the thickness exceeds the optimal value, the decrease in prompt efficiency becomes dominant and this lessens the detection efficiency. This optimal value is determined to be around 50-60 $\mu$m for all designs. Secondly, the average neutron capture time of each detector design is obtained. With the mean capture time of 56 $\mu$s, our proposed design HSP reduces the mean capture time by about $8\%$ according to the Panda design in spite of having lower Gd concentration (0.18 vs 0.19). Thirdly, the variation of neutron capture efficiency of the detectors depending on the correlated time window is searched. It is revealed that the time required to attain efficiency saturation significantly changes with respect to the detector type. However, it is seen that as the correlated time window is expanded, the effect of detector design on the neutron capture efficiency decreases. Finally, the detection efficiency of each detector depending on the correlated time window is calculated when the prompt and delayed energy threshold is applied. The required minimum time window to reach the maximum efficiency is determined for each detector type. The best result is obtained with the Panda2 since it achieves the highest efficiency within the shortest time.

Additionally, the light collection efficiency of the detectors is compared. It is found that Panda, Panda2, and HSP collect almost the same amount of light ($18\%$ for 2-inch and $35\%$ for 3-inch PMT ), while the Cormorad collects only half of the light ($9\%$ for 2-inch and $18\%$ for 3-inch) collected by other detectors. As a result, the best performance is obtained with the HSP design since it achieves the same efficiency by using $9\%$ fewer PMTs compared to Panda and Panda2 (182 vs 200). In this manner, efficient use of PMTs comes from the honeycomb structure of HSP. Constructing the HSP detector in a honeycomb form minimizes the surface areas of the bars and thus reduces the number of PMTs required to readout a given detector volume. Considering all of these findings, the proposed design HSP may be a good alternative to conventional design for future segmented antineutrino detector construction.

\clearpage




\bibliographystyle{elsarticle-num}

\bibliography{sample}

\end{document}